\documentclass[twocolumn, iop]{emulateapj}
\usepackage{apjfonts}
\usepackage{color}

\newcommand{\RNum}[1]{\uppercase\expandafter{\romannumeral #1\relax}}

\usepackage{amsmath,graphicx,longtable,hyperref}
\usepackage{natbib,threeparttable,rotating}
\usepackage{ulem}

\usepackage[title]{appendix}
\usepackage{subfigure}
\usepackage{amsmath}
\usepackage{multirow}

\newcommand{\etal}{et al.}
\newcommand{\hbeta}{H{$\beta$}}
\newcommand{\halpha}{H{$\alpha$}}

\newcommand{\CIV}{C\,{\sevenrm IV}}

\def\FeII{Fe\,{\sc ii}}
\def\MgII{Mg\,{\sc ii}}
\def\HeII{He\,{\sc ii}}

\def \OIII {[O\,{\sc iii}]}

\newcommand{\NII}{[N\,{\sevenrm\,II}]}

\newcommand{\SII}{[S{\sevenrm\,II}]}

   \font\sevenrm=cmr7 scaled 1000
\newcommand{\comments}[1]{}

\def\kms{{\rm km\,s^{-1}}}

\begin{document}

\title{The Sloan Digital Sky Survey Reverberation Mapping Project: Low-Ionization Broad-line Widths and Implications for Virial Black Hole Mass Estimation}

\author{Shu Wang$^{1,2}$, Yue Shen$^{3,4,*}$,  Linhua Jiang$^{1,2}$, Keith Horne$^5$, W.~N. Brandt$^{6,7,8}$, C.~J. Grier$^{6,9}$, Luis C.~Ho$^{1,2}$, Yasaman Homayouni$^{10}$, Jennifer I-Hsiu Li$^{3}$, Donald P. Schneider$^{6,7}$, Jonathan R.~Trump$^{10}$} 

\altaffiltext{1}{Kavli Institute for Astronomy and Astrophysics, Peking University, Beijing 100871, China; wangshukiaa@pku.edu.cn, jiangKIAA@pku.edu.cn}
\altaffiltext{2}{Department of Astronomy, School of Physics, Peking University, Beijing 100871, China}
\altaffiltext{3}{Department of Astronomy, University of Illinois at Urbana-Champaign, Urbana, IL 61801, USA}
\altaffiltext{4}{National Center for Supercomputing Applications, University of Illinois at Urbana-Champaign, Urbana, IL 61801, USA}
\altaffiltext{5}{SUPA Physics and Astronomy, University of St. Andrews, Fife, KY16 9SS, Scotland, UK}
\altaffiltext{6}{Department of Astronomy \& Astrophysics, The Pennsylvania State University, University Park, PA, 16802, USA}
\altaffiltext{7}{Institute for Gravitation and the Cosmos, The Pennsylvania State University, University Park, PA 16802, USA}
\altaffiltext{8}{Department of Physics, 104 Davey Lab, The Pennsylvania State University, University Park, PA 16802, USA }
\altaffiltext{9}{Steward Observatory, University of Arizona, 933 North Cherry Avenue, Tucson, AZ 85721-0065, USA}
\altaffiltext{10}{University of Connecticut, Department of Physics, 2152 Hillside Road, Unit 3046, Storrs, CT 06269-3046, USA}
\altaffiltext{$^*$}{Alfred P. Sloan Research Fellow.}

\shorttitle{SDSS-RM: Quasar Broad-Line Widths}
\shortauthors{Wang \etal}

\begin{abstract}

The width of the broad emission lines in quasars is commonly characterized either by the full-width-at-half-maximum (FWHM) or the square root of the second moment of the line profile ($\sigma_{\rm line}$), and used as an indicator of the virial velocity of the broad-line region (BLR) in the estimation of black hole (BH) mass. We measure FWHM and $\sigma_{\rm line}$ for \halpha, \hbeta\ and \MgII\ broad lines in both the mean and root-mean-square (rms) spectra of a large sample of quasars from the Sloan Digital Sky Survey Reverberation Mapping (SDSS-RM) project. We introduce a new quantitative recipe to measure $\sigma_{\rm line}$ that is reproducible, less susceptible to noise and blending in the wings, and scales with the intrinsic width of the line. We compare the four definitions of line width (FWHM and $\sigma_{\rm line}$ in mean and rms spectra, respectively) for each of the three broad lines and among different lines. There are strong correlations among different width definitions for each line, providing justification for using the line width measured in single-epoch spectroscopy as a virial velocity indicator. There are also strong correlations among different lines, suggesting alternative lines to \hbeta\ can be used to estimate virial BH masses. We further investigate the correlations between virial BH masses using different line width definitions and the stellar velocity dispersion of the host galaxies, and the dependence of line shape (characterized by the ratio FWHM/$\sigma_{\rm line}$) on physical properties of the quasar. Our results provide further evidence that FWHM is more sensitive to the orientation of a flattened BLR geometry than $\sigma_{\rm line}$, but the overall comparison between the virial BH mass and host stellar velocity dispersion does not provide conclusive evidence that one particular width definition is significantly better than the others.
\end{abstract}

%We provide the best fitting results of these correlations and quantify their intrinsic scatter, which can be used to inter-calibrate recipes for virial BH mass estimation using different lines. 

\section{Introduction}\label{sec:intro}

Measuring BH masses in quasars is of critical importance to most studies of supermassive black holes (SMBHs). The primary method to measure BH masses is the reverberation mapping (RM) \citep[e.g.,][]{Blandford_McKee_1982,Peterson_1993}, where the time lag $\tau$ between the continuum and broad emission line variability provides a measurement of the characteristic size of the BLR, i.e., $R=c\,\tau$. Assuming the BLR is virialized, a virial BH mass can be computed using the broad-line velocity width $W$ as an indicator of the virial velocity: $M_{\rm BH}=f\frac{W^2R}{G}$, where $f$ is the dimensionless virial coefficient that accounts for the geometry and kinematics of the BLR. Local RM results have discovered a remarkable correlation between the size of the BLR and the luminosity of the AGN, known as the $R-L$ relation \citep[e.g.,][]{kaspi_reverberation_2000, bentz_low-luminosity_2013}, which was subsequently utilized to develop the single-epoch virial BH mass estimators that allow the estimation of a BH mass using a single spectrum \citep[e.g.,][]{Vestergaard_Peterson_2006,shen_mass_2013}.  

The only directly measured quantity used in the virial BH mass estimation is the width of the broad line. There are two commonly used characterizations of line width \citep[e.g.,][]{peterson_central_2004}, the FWHM and the square root of the second moment of the line (referred to as the line dispersion $\sigma_{\rm line}$). For single-epoch spectroscopy, both FWHM and $\sigma_{\rm line}$ are measured from the spectrum. However, for objects with reverberation mapping spectroscopy, both line widths can also be measured from the rms spectrum, which better represents the variable component of the line. Although there are correlations between these different line width definitions \citep[e.g.,][hereafter, C06]{collin_systematic_2006}, it remains a topic of debate as to which line width definition is better to indicate the virial velocity of the BLR. The pros and cons of these different line width definitions have been discussed at length in, e.g., \citet{peterson_central_2004, peterson_masses_2011}. From a practical point of view, FWHM is relatively easier to measure and less susceptible to blending and noisy wings of the line than $\sigma_{\rm line}$ \citep[e.g.][]{peterson_central_2004}; however, FWHM is more sensitive to inappropriate narrow-line removal, and may be more affected by orientation in a flattened geometry of the BLR than $\sigma_{\rm line}$ \citep[e.g.,][]{Wills_Browne_1986,Runnoe_etal_2013a,Shen_Ho_2014,Brotherton_etal_2015,Mejia-Restrepo_etal_2018a}. On the other hand, there may be systematic differences in the line widths measured from the mean and rms spectra. 

C06 performed the first systematic comparison among the four types of line width measurements (FWHM and $\sigma_{\rm line}$ from both mean and rms spectra) using 35 low-redshift AGN with RM measurements. By comparing the virial products based on the four different width measurements with the stellar velocity dispersion \citep[used as an independent indicator of the true BH mass via the $M_{\rm BH}-\sigma_*$ relation, e.g.,][]{Tremaine_etal_2002} of 14 host galaxies, they concluded that $\sigma_{\rm line}$ is a better indicator of the virial velocity than FWHM, as the latter displayed a larger variation of the virial coefficient $f$ across sub-populations of objects. 

In this work we expand the C06 with a new RM sample from the SDSS-RM project \citep[][]{Shen_etal_2015a}. Our sample is the largest to date with which one can measure all four types of broad-line widths for multiple broad lines.  In \S\ref{sec:sample} we describe the sample and the data used and in \S\ref{sec:tech} detail our methodology for obtaining line width measurements. We present our main results in \S\ref{sec:results}, discuss their significance in \S\ref{sec:disc} and conclude in \S\ref{sec:con} with a discussion on future perspectives.

\section{Data and Sample}\label{sec:sample}

SDSS-RM is a dedicated RM project that uses the SDSS BOSS spectrograph \citep[][]{Smee_etal_2013} on the 2.5m SDSS telescope \citep{gunn_2.5_2006} to monitor 849 broad-line quasars in a single $7\,{\rm deg^2}$ field over a broad redshift and luminosity range \citep{Shen_etal_2015a}. The SDSS-RM sample is a flux-limited sample, and is designed to be a representative sample of the general quasar population without any cuts on spectral and variability properties of quasars. A detailed description of the sample characterization is presented in \citet{shen_sample_2019}. From the commencement of this program as part of SDSS-III \citep{eisenstein_sdss-iii:_2011}, SDSS-RM obtained 32 spectroscopic epochs in 2014 at an average cadence of $\sim 4$ days and will continue through 2020 with 6-12 spectroscopic epochs per observing season.

The spectroscopic data used in this work is the 32 epochs taken during 2014. Both the individual epoch spectra and the coadded spectra during this period are used in our analysis. The wavelength coverage of BOSS spectrographs is $\sim3650-10400\textrm{\AA}$, with a spectral resolution of $R\sim2000$. The spectroscopic data is first pipeline-processed as part of the SDSS-III Data Release 12 \citep{alam_eleventh_2015} followed by a custom flux calibration scheme and improved sky subtraction as described in \cite{Shen_etal_2015a}. The improved spectrophotometry has a nominal accuracy of ~5\%.

In order to improve further the flux calibration and to construct rms spectra from the first-year data, we use a custom program called ``PrepSpec'' developed by K. Horne. As detailed in \citet{shen_sloan_2016-1}, PrepSpec models the time resolved spectroscopic data set with a separable model that accounts for the variations in the broad line and continuum flux. During this process, the fluxes of the narrow emission lines, assumed to be constant during the monitoring period, are used to adjust and improve the overall flux calibration. PrepSpec measures the continuum and broad line fluxes as time series from the model, and generates the rms spectra for the major broad emission lines. This approach differs from earlier RM work \citep[e.g.,][]{peterson_central_2004}; the construction of the rms spectra isolates the continuum variability from the broad-line variability, and traces the variability in the broad lines. PrepSpec employs rigorous statistical modeling (assuming Gaussian errors) to create the light curves (continuum and broad-line variability) and the rms line profile along with their measurement uncertainties. These rms broad-line spectra are used to measure the broad-line widths in our study.

\section{Line width measurement}\label{sec:tech}

This work compares the two commonly adopted line width definitions, FWHM and $\sigma_{\rm line}$, for different broad lines. Since both FWHM and $\sigma_{\rm line}$ can be measured from the mean and rms spectra, we consider four different line width measurements: FWHM$_{\rm mean}$, FWHM$_{\rm rms}$, $\sigma_{\rm line, mean}$ and $\sigma_{\rm line, rms}$.

\subsection{Mean and rms spectra }\label{sec:spec} 

The mean spectra are constructed by coadding all 32 epochs for each of the 849 quasars in the SDSS-RM sample using the SDSS-III spectroscopic pipeline idlspec2d.\footnote{Publicly available at http://www.sdss3.org/svn/repo/idlspec2d/} These coadded spectra are nearly identical to those generated with a simple arithmetic mean or inverse-variance-weighted mean; therefore, this detail does not significantly affect the measurements of the broad-line widths. PrepSpec also outputs a mean spectrum for each quasar that is essentially the same as our own coadded spectrum. 

There are two approaches to construct the rms broad-line spectrum. The commonly used method \citep[e.g.,][]{peterson_central_2004} is to directly compute the rms value, pixel-by-pixel, over all epochs for the total spectral flux (continuum plus lines); the rms line profile is obtained by subtracting the local continuum in the total rms spectrum. The second approach is to subtract the continuum in each epoch and compute the rms from the line-only spectra. As noted by \cite{barth_lick_2015}, there are systematic differences between the line-only rms spectra from these two approaches. Appendix A provides our simulations to demonstrate such differences. The main reason that earlier RM studies adopted the former approach is that it is relatively straightforward to compute and does not rely on specific modeling of the decomposition of continuum and line flux, although the resultant rms line spectrum generated is a biased measure of the true line variability. PrepSpec generates the rms line spectrum following the second approach, where the rms variability is from the broad emission line only.

\subsection{Spectral fitting of the mean spectrum}\label{sec:fitmethod}

\begin{figure}[h] 
\begin{center}
\includegraphics[width=0.48\textwidth]{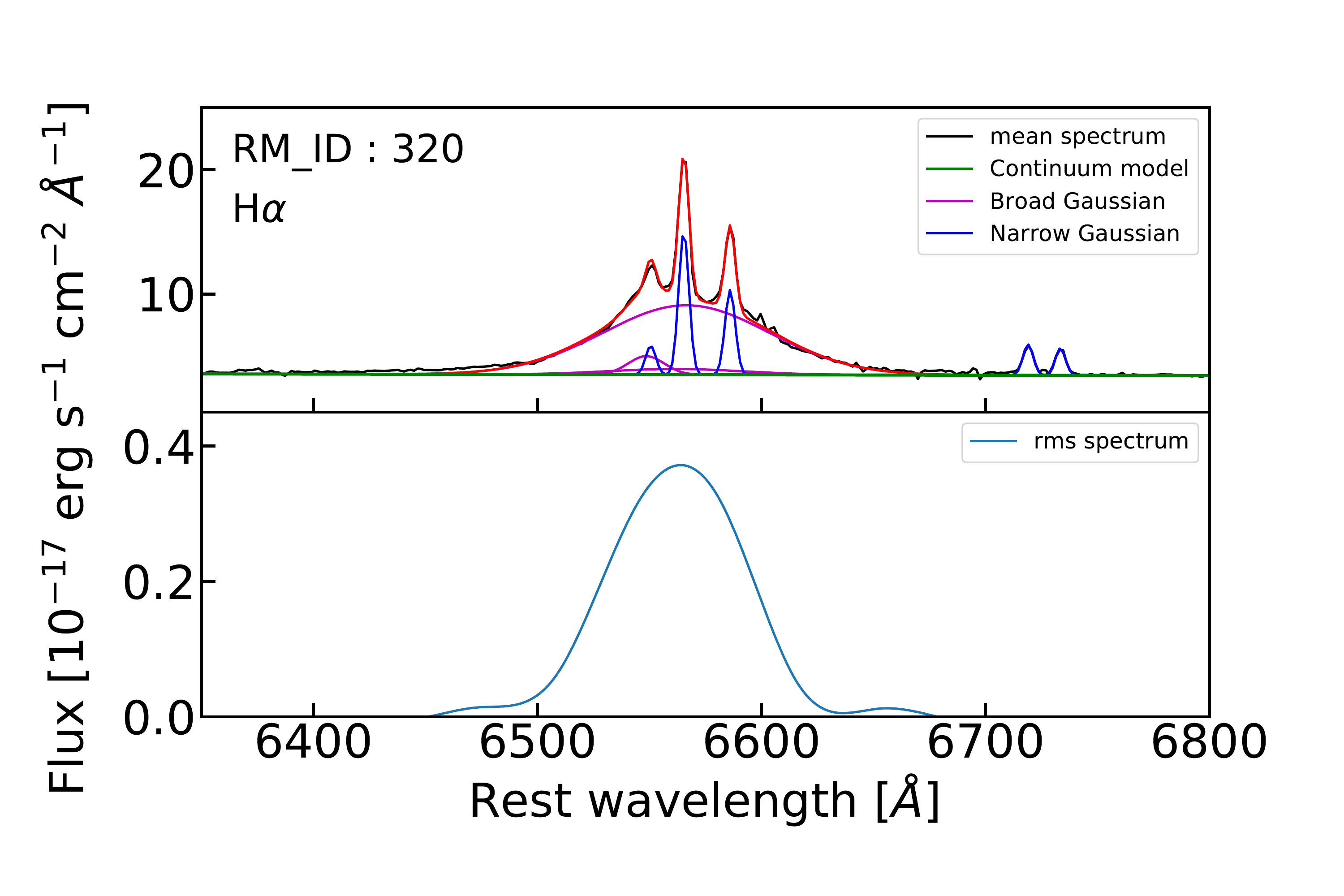} 
\includegraphics[width=0.48\textwidth]{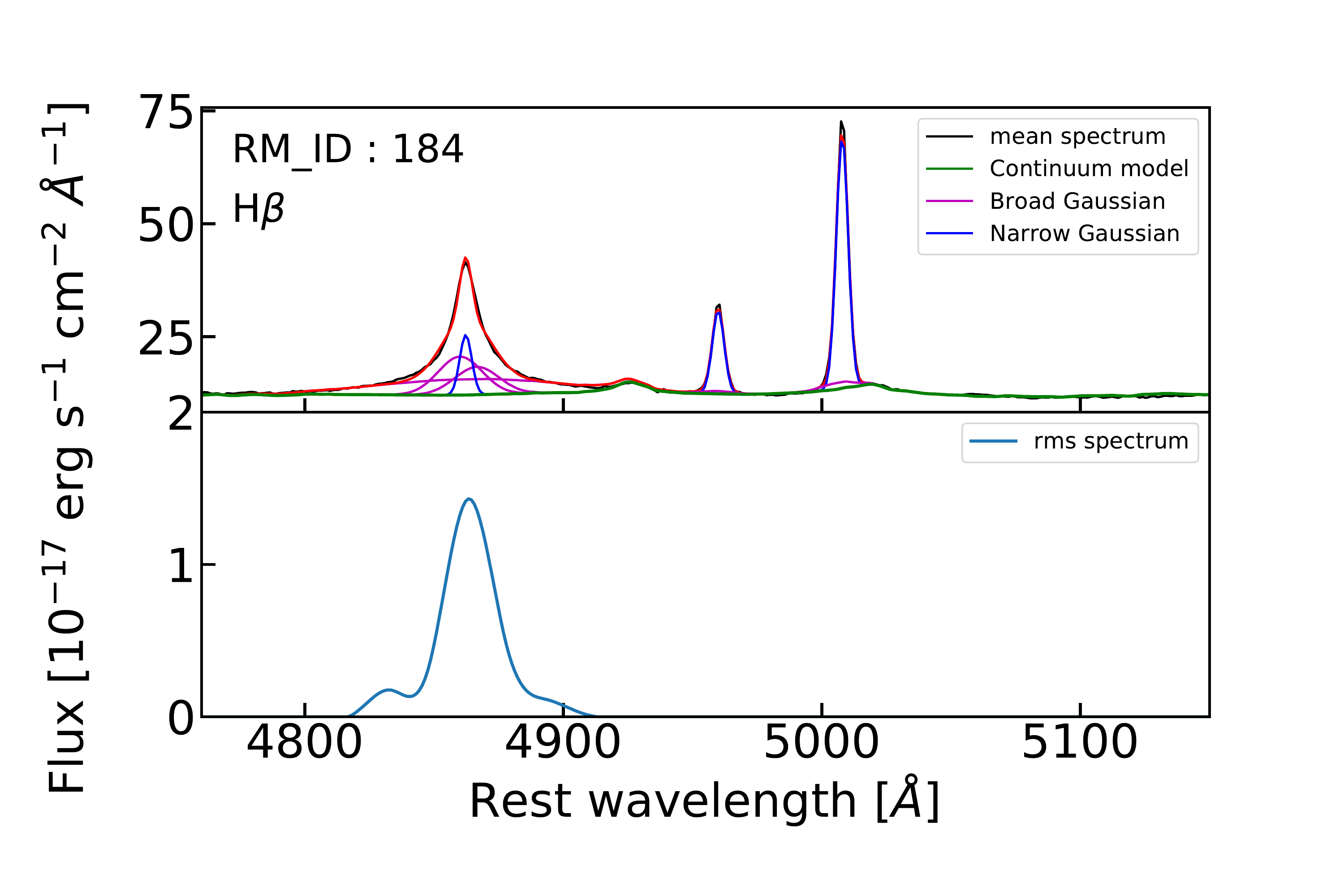} 
\includegraphics[width=0.48\textwidth]{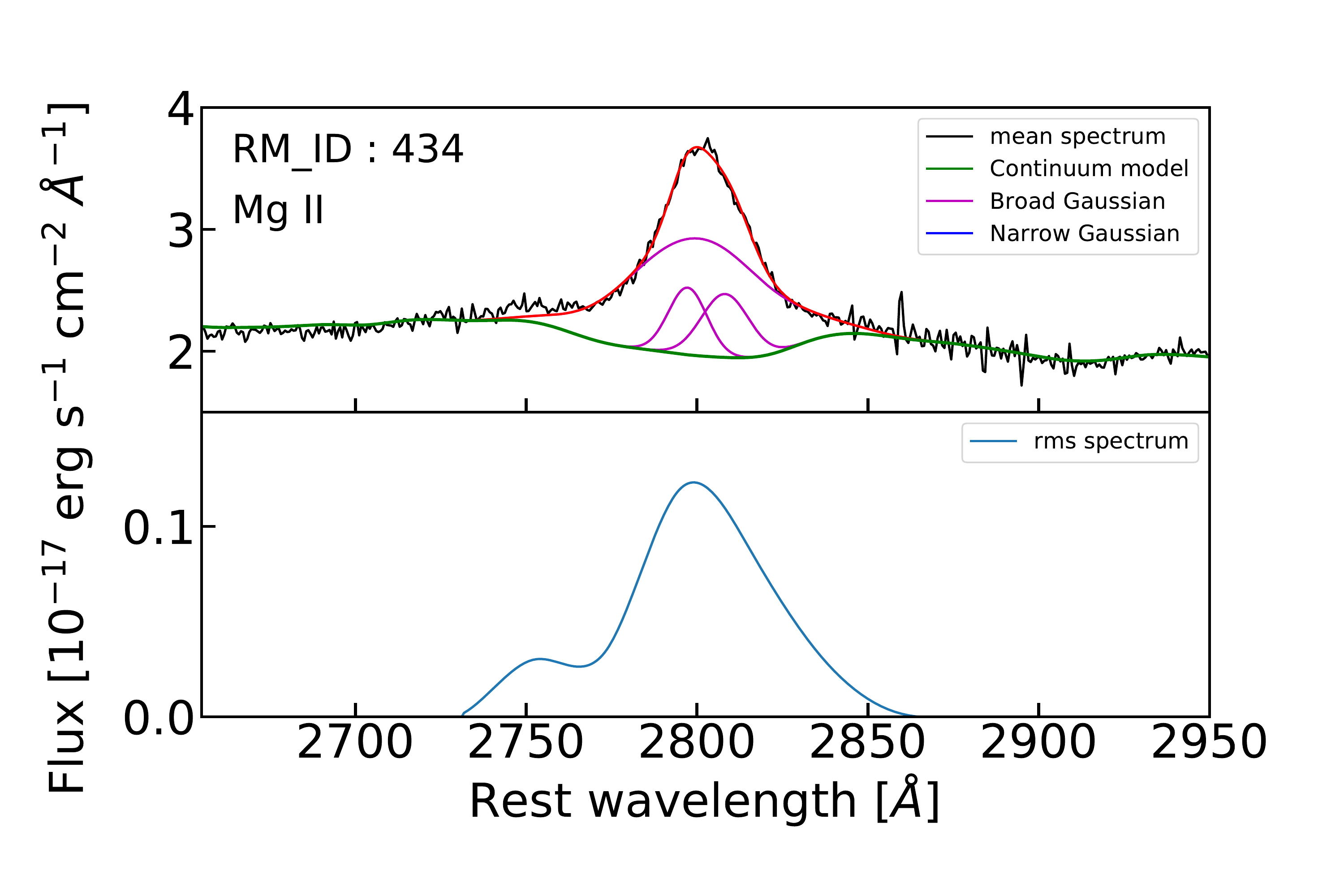} 
\caption{Examples of line fitting results from the mean and rms spectra for \halpha, \hbeta\ and \MgII\ from top to bottom, respectively. The RMID of each object is shown in the upper-left corner. Black lines are the mean spectra. Green lines are the continuum model which is the sum of the power-law component, \FeII\ emission, and the polynomial component (for \MgII\ only).  Magenta and blue lines indicate each of the broad and narrow Gaussian components, respectively. Red lines represent the total line model. All line fitting results in the figure are calculated by adding the underlying continuum model to match the mean spectrum. The rms spectrum from PrepSpec is shown in the bottom of each panel as the blue line.  } \label{fig:haexample} 
\end{center}
\end{figure}

The multi-component functional fitting method is used to decompose the mean spectra following our earlier work \citep[e.g.,][]{shen_biases_2008, shen_catalog_2011}. We use the publicly available code {\tt QSOFIT} \citep{shen_sample_2019} with small modifications to perform our spectral fits. Examples of our fitting result are displayed in Figure \ref{fig:haexample}. Compared with direct pixel measurement from the original spectrum, this approach is more robust against noise and artifacts in the reduction process. To confirm that the model accurately reproduces the data, we visually inspect all the fitting results and reject poorly-fit objects from later analysis. Rejection using a cut on the reduced $\chi^2$ statistic results in a similar subsample as our visual inspection. 

In this work, we focus on three broad emission lines: \halpha, \hbeta\ and \MgII. These lines are covered in the low-redshift and low-luminosity subset of the SDSS-RM sample, and therefore their variations are well captured by the 2014 spectroscopy. High-ionization lines, such as \CIV, are present in the high-redshift subset of the SDSS-RM sample. The 2014 data does not cover a sufficient time baseline to probe the \CIV\ variability for most objects in our sample. An extension of the current work to \CIV\ and other broad lines covered in the high-$z$ SDSS-RM subsample will be presented in future work using the final multi-year SDSS-RM data set. 

We first fit the continuum and broad-band \FeII\ emission near the broad emission line of interest and subtract the pseudo-continuum (continuum plus \FeII\ emission) from the spectrum, leaving a line-only residual spectrum (see below). Each broad line in the residual spectrum is fit using multiple Gaussians. We chose to fit each line locally to avoid complex continuum shapes over broad spectral ranges.  In the regions adjacent to either \halpha\ or \hbeta, emission lines existing within each line complex are fit simultaneously with Gaussian functions. Table \ref{table:fitparameter} lists the names, central wavelengths, and the number of Gaussians for each line component in the fits.  

To fit the local continuum, we choose windows that are free of prominent emission or absorption lines. The specific continuum windows for each line complex are described below. The continuum model consists of a power law and \FeII\ emission constrained from templates. In the continuum fitting of \MgII, a low-order polynomial in $\lambda$ is added in order to account for possible reddening in the rest-frame UV regime. The \FeII\ emission is modeled using templates from \cite{boroson_emission-line_1992} in the optical and \cite{vestergaard_empirical_2001} in the UV.

To fit each line complex, we transform the spectrum into velocity space using the vacuum wavelength of the line and the redshift of the quasar. As shown in Table \ref{table:fitparameter}, the broad lines are fit with 3 Gaussians and the narrow lines are fit with 1 Gaussian in each complex. The only exception is \OIII\ $\lambda\lambda$ 4959, 5007 where one additional Gaussian is added to account for the blue asymmetric wing \citep[e.g.,][]{peterson_observations_1981,  Shen_Ho_2014}. The broad and narrow line division is set to be 400 km s$^{-1}$ in Gaussian $\sigma$, which is $\sim$940 $\kms$ in FWHM. This division works well for \halpha\ and \hbeta\ in all objects in our sample. During the fitting, all narrow line centers and widths are tied together in each line complex, and the flux ratios of the doublets are set to their theoretical values. In most cases, the initial estimate of the line center is set to zero velocity. The initial width of the narrow lines is set to $\sigma=250\,\kms$,  while the initial width of the broad lines is set to $\sigma=650$ km s$^{-1}$ for two Gaussians and 1500 km s$^{-1}$ for the the third Gaussian. 

The specific fitting details for each line complex are:
\begin{itemize} 
\item[$\bullet$] \halpha: The continuum windows are [6150, 6250]\AA\ and [6800,7000]\AA. After subtracting the continuum model, the \SII\ $\lambda\lambda$\,6718,6732 doublet is fitted. The reason for this step is to obtain a prior on the center and especially the width of the narrow lines to help decompose the narrow \halpha\ component more accurately. The \SII\ doublet is fitted using three Gaussians, two for the doublet and one additional Gaussian as an approximation of the red wing of broad \halpha. All lines in the complex are fitted simultaneously, including broad and narrow \halpha, \NII\ $\lambda\lambda$ 6549, 6585 doublet, and \SII\ $\lambda\lambda$ 6718, 6732 doublet. If the fit of \SII\ is acceptable (i.e., reduced $\chi^{2}$ less than 10), the values of line centers and widths for all narrow line components (narrow \halpha\, \NII\ $\lambda\lambda$ 6549, 6585) are tied to the values of \SII. If the \SII\ fit has a reduced $\chi^2$ greater than 10, the initial values of all narrow lines are set as the value described above. Our spectra generally have very high S/N per pixel hence our Gaussian fit may not be a good fit to the \SII\ line in a statistical sense. Therefore we use a moderately large reduced $\chi^2$ cut to exclude obviously bad \SII\ fits. The line flux ratio of the \SII\ doublet is fixed to 1 and that of \NII\ $\lambda$ 6585 to \NII\ $\lambda$ 6549 is fixed to 3.

\item[$\bullet$] \hbeta: The continuum windows are [4450, 4630]\AA, [4750, 4770]\AA\ and [5050, 5500]\AA. These windows are chosen to avoid contamination by H$\gamma$ and \HeII\ $\lambda\;4686$. After subtracting the continuum model, all lines are simultaneously fitted including broad and narrow \hbeta\ as well as the \OIII\ doublet. Each line in the \OIII\ doublet is fitted with two Gaussians, one for the core component and the other for the blueshifted wing component. For the wing Gaussian, the initial center is set to 200 km s$^{-1}$ blueward of the \OIII\ vacuum wavelength and the initial $\sigma$ of the Gaussian is set to 510 km s$^{-1}$. The flux ratio of the \OIII\ doublet is not constrained to allow better fitting for strong \OIII\ lines.

\item[$\bullet$] \MgII: The continuum windows are [2300, 2700]\AA\ and [2900, 3400]\AA. An additional 3rd-order polynomial in $\lambda$ is added to account for possible reddening seen in some objects \citep[e.g.,][]{shen_sample_2019}. In the line fitting, \MgII\ is not treated as a doublet because the line is generally too broad to separate the doublet. The initial estimates and limits on the fitting parameters are set as the fiducial values described above.

\end{itemize}

\begin{table}  
\caption {Line fitting parameters} \label{table:fitparameter}
\begin{center}
\begin{tabular}{c c c c  } 
\hline \hline
Line  & Rest wavelength$^a$ (\AA) & N${_{gauss}}$ & Complex \\ \hline
\halpha & 6564.61 & 4$^b$ & \halpha  \\  
\SII  & 6732.67 & 1 & \halpha  \\  
\SII & 6718.29 & 1 & \halpha  \\  
\NII & 6549.85 & 1 & \halpha  \\ 
\NII & 6585.28 & 1 & \halpha  \\  
\hbeta & 4862.68 & 4$^b$ & \hbeta  \\  
\OIII &  4960.30  & 2 & \hbeta  \\  
\OIII &  5008.24  & 2 & \hbeta \\  
\MgII &  2798.75  & 4$^b$ & \MgII \\  \hline 
 \multicolumn{4}{p{0.45\textwidth}}{Notes. a. The rest wavelengths are in vacuum. b. Three Gaussians are used to fit the broad-line component and one Gaussian is used for the narrow-line component.}
\end{tabular}
\end{center}
\end{table}

\subsection{Line width measurements}

While PrepSpec also outputs broad-line widths from the mean and rms spectra, we decided to perform our own measurements. This is because the current version of PrepSpec does not perform a spectral decomposition in the mean spectrum as sophisticated as our approach described above and some line widths reported by PrepSpec are biased. In addition, we will also investigate different choices of windows in measuring $\sigma_{\rm line}$, requiring our own analysis. Visual inspection of the spectral modeling and line width measurements for individual objects suggests that our own line width measurements are generally more reliable than the default PrepSpec outputs. 

The sum of all broad Gaussians in our spectral decomposition of the mean spectrum is used to measure FWHM$_{\rm mean}$ and $\sigma_{\rm line, mean}$, and the rms model spectra from PrepSpec are used for the measurements of FWHM$_{\rm rms}$ and $\sigma_{\rm line, rms}$.

To measure FWHM, we first locate the peak of the line model, and identify the half-peak positions on both the red and blue sides of the peak, $\lambda_{\rm red half}$ and $ \lambda_{\rm blue half}$. The FWHM is calculated as 
\begin{equation} \label{FWHMcalc}
{\rm FWHM}=\lambda_{\rm red half} - \lambda_{\rm blue half}.
\end{equation}

\noindent $\sigma_{\rm line}$ is calculated from its definition \citep[e.g.,][]{peterson_central_2004}:
\begin{equation} \label{eqsigma}
\sigma_{\rm line}^{2} = \langle \lambda^{2} \rangle - \lambda_{0}^{2} = \left( \int \lambda^{2} F(\lambda) d\lambda \right) \;/ \left( \int F(\lambda) d\lambda \right) - \lambda_{0}^{2},
\end{equation}

\noindent where $F(\lambda)$ is the line profile and $\lambda_{0}$ is the first moment of $F(\lambda)$: 

\begin{equation} \label{eqlambda0}
\lambda_{0} = \left( \int \lambda F(\lambda) d\lambda \right) \; / \left( \int F(\lambda) d\lambda \right).
\end{equation}

\noindent After that, we convert FWHM and $\sigma_{\rm line}$ to velocity units. 

In equation (\ref{eqsigma}) the integrand in the first term is proportional to $\lambda^{2}$, which means the $\sigma_{\rm line}$ measurement is sensitive to window choices. Ideally the integration in equation (\ref{eqsigma}) should be over the entire wavelength range of the broad line. However, in practice this approach is not possible because the broad line profile is typically poorly constrained in the wings due to the limited S/N of the spectrum. To make our $\sigma_{\rm line}$ measurement repeatable, we define a window that can be easily computed for any spectrum. We tested using multiples of FWHM$_{\rm mean}$ and MAD$_{\rm mean}$ (the Mean Absolute Deviation) as the window.  MAD is defined as: 

\begin{equation} \label{eqmed}
{\rm MAD} = \left( \int \mid \lambda- \;{\rm MED}\; \mid  \;F(\lambda) d\lambda \right) \; / \left(\int F(\lambda) d\lambda \right),
\end{equation}

\noindent where MED is the median wavelength of the line profile, and is defined as the location where the integrated flux (weight) from the blue side to this location is half of the total line flux. 

We finally adopt 2.5$\times$MAD$_{\rm mean}$ as the {\it half-window size} to calculate $\sigma_{\rm line}$ for both mean and rms spectra. For mean spectra, the window is centered at the vacuum wavelength of the line. For rms spectra the line center is re-defined as the peak of the rms profile in order to enclose much of the rms flux, since the rms profile can be significantly asymmetric. This choice of a MAD$_{\rm mean}$-based window has several advantages: it is a well-defined window that can be reproduced in other work; MAD itself is a measure of the line width, so the window is automatically adjusted according to line width; using such a window to calculate $\sigma_{\rm line}$ mitigates noise or artifacts in the wings of the line, as well as contamination from residuals from the adjacent narrow line (or blended line) removal. Appendix B provides a detailed discussion on our window choice.

We employed the Monte Carlo approach described in \cite{shen_biases_2008, shen_catalog_2011} to estimate uncertainties in the line width measurements. To create a mock spectrum, the original spectrum is perturbed at each pixel by a random deviation drawn from a Gaussian distribution whose $\sigma$ is set to the flux density uncertainty at that pixel. After generating a mock spectrum, we apply the same fitting approach to derive the spectral measurements. We generate 50 mock spectra for each object and estimate the measurement uncertainty as the semi-amplitude of the range enclosing the 16th and 84th percentiles of the distribution.

\subsection{Subsamples with good quality}

In order to mitigate the effects of low-quality measurements, we select subsamples with good quality in their fitting results, which are designated as the ``good'' samples. For mean-spectrum line width measurements, we select a high signal-to-noise subsample using equation (\ref{eqcrmean1}) below:

\begin{equation}
\{S/N\}_{\rm line, mean} >3 \label{eqcrmean1},
\end{equation}

\noindent where $\{S/N\}_{\rm line, mean}$ is the average flux to uncertainty ratio in the total-flux mean spectrum within the line-fitting window. 

%Equation (\ref{eqcrmean2}) below is used to reject bad fitting results:
%\begin{equation}
%\chi_{line fit}^2/dof < 10 \;\;\;\; \chi_{continuum\ fit}^2/dof < 10 \label{eqcrmean2} \\
%\end{equation}
%
%\noindent where $\chi_{\rm line fit}^2$ is the normalized sum of squared deviations between model and data; $dof$ is the degree of freedom in the fitting. 

Two additional criteria, equation (\ref{eqcrmean3}) and (\ref{eqcrmean4}) are imposed to require that the mean line widths are well measured:
\begin{gather}
{\rm FWHM}_{\rm mean}\;/\; {\rm FWHM}_{\rm mean,error}\; > 3 \label{eqcrmean3},\\
\sigma_{\rm line, mean}\;/\; \sigma_{\rm line,mean,error} > 3 \label{eqcrmean4}.
\end{gather}

In addition to the above quantitative criteria, we also visually inspected the fits to the mean spectra and excluded objects where the spectrum is heavily affected by skylines, has moderate to strong absorption features, covers less than half of the profile, or the model clearly failed to account for the complex profile. In total, 11 \hbeta, 6 \halpha\ and 67 \MgII\ cases are excluded from our visual inspection.

For rms-spectrum line width measurements, we first select a high-variability subsample using equation (\ref{eqcrrms1}):
\begin{equation}
\{S/N\}_{\rm lightcurve} = \sqrt{{\chi_{\rm lightcurve}^{2} - (N_{\rm epoch}-1)}} > 10 \label{eqcrrms1},
\end{equation}

\noindent {where $\{S/N\}_{\rm lightcurve}$ quantifies the intrinsic variability of the broad-line light curve \citep[equivalent to ``SNR2'' in the sample catalog compiled by][]{shen_sample_2019}. We set this criteria to ensure that the rms line profile is well determined by PrepSpec and dominated by intrinsic broad emission line variability. As for the mean line widths, equation \ref{eqcrrms2} and equation \ref{eqcrrms3} are imposed to require that the rms line widths are well measured:

\begin{gather}
{\rm FWHM}_{\rm rms}\;/\; {\rm FWHM}_{\rm rms,error}\; > 3 \label{eqcrrms2}, \\
\sigma_{\rm line, rms}\;/\; \sigma_{\rm line,rms,error} > 3 \label{eqcrrms3}.
\end{gather}

For comparisons involving only mean line widths, the criteria are equations\ (\ref{eqcrmean1}-\ref{eqcrmean4}), while the criteria are equations\ (\ref{eqcrrms1}-\ref{eqcrrms3}) for comparisons only involving rms line widths. For comparisons involving both mean and rms line widths, all criteria are included. The number of objects that cover each line and the number of objects in the good subsamples that pass the criteria are listed in Table \ref{table:subsamplesize}. The measured line widths with all four definitions for \halpha, \hbeta\ and \MgII, along with additional fitting parameters are provided in an online fits table; its content is summarized in Table \ref{table:finaltable}.

\begin{table}
\caption {Number of objects in each subsample} \label{table:subsamplesize}
\begin{center}
%\begin{tabular}{p{1.8cm}<{\centering}  p{0.8cm}<{\centering}  p{1.2cm}<{\centering}  p{1.2cm}<{\centering}  p{2cm}<{\centering} } 
\begin{tabular}{c c c c c}
\hline \hline
                                & Total        & Good mean	& Good rms 	& Good mean and rms$^a$\\ \hline
\halpha\ 			& 58   	&  45 		& 39			& 38 \\
\hbeta\                      & 222        & 170		& 81			& 76 \\	
\MgII\ 			& 755	& 512		& 329		& 296 \\ 
\halpha\ \& \hbeta\ 	& 58 		& 43  		& 31			& - \\
\MgII\ \& \hbeta\ 	& 204  	& 114		& 43			& -\\
\hline
 \multicolumn{5}{p{0.45\textwidth}}{Notes. a. The numbers of the good mean and rms sample for \halpha\ \& \hbeta\ and \MgII\ \& \hbeta\ are not shown since we did not use such samples in our study. }
\end{tabular}
\end{center}
\end{table}

\begin{table*}
\caption {Line width measurements} \label{table:finaltable}
\begin{center}
        \begin{tabular}{c c c c p{9cm} } 
        \hline  \hline
        Column No. &  Column name & Type & {Units} & Description \\ \hline
        0	&  RMID & INT & - &SDSS-RM Identification number\\
        1       & SDSS name & STRING & - & Name of the object in SDSS \\
        2	& RA & DOUBLE & degree & Right Ascension (J2000) \\
        3	& DEC & DOUBLE & degree & Declination (J2000) \\ 
        4       & Redshift & DOUBLE & - & Redshift \\
        5       & SIGMA$\_$HOST   & DOUBLE[2]   & km s$^{-1}$ & Stellar velocity dispersion and its measurement uncertainty \\
	6	& HALPHA$\_$FWHM$\_$mean  & DOUBLE[2] & km s$^{-1}$ & \halpha\ FWHM (and measurement uncertainty) in mean spectra\\	
	7	& HALPHA$\_$SIGMA$\_$mean & DOUBLE[2] & km s$^{-1}$ & \halpha\ $\sigma_{\rm line}$ (and measurement uncertainty) in mean spectra \\
	8	& HALPHA$\_$FWHM$\_$rms & DOUBLE[2] & km s$^{-1}$ & \halpha\ FWHM (and measurement uncertainty) in rms spectra \\
	9	& HALPHA$\_$SIGMA$\_$rms & DOUBLE[2] & km s$^{-1}$ & \halpha\ $\sigma_{\rm line}$ (and measurement uncertainty) in rms spectra \\ 
	10	& HBETA$\_$FWHM$\_$mean  & DOUBLE[2] & km s$^{-1}$ & \hbeta\ FWHM (and measurement uncertainty) in mean spectra\\	
	11	& HBETA$\_$SIGMA$\_$mean & DOUBLE[2] & km s$^{-1}$  & \hbeta\ $\sigma_{\rm line}$ (and measurement uncertainty) in mean spectra \\
	12	&  HBETA$\_$FWHM$\_$rms & DOUBLE[2] & km s$^{-1}$ & \hbeta\ FWHM (and measurement uncertainty) in rms spectra \\
	13	& HBETA$\_$SIGMA$\_$rms & DOUBLE[2] & km s$^{-1}$ & \hbeta\ $\sigma_{\rm line}$ (and measurement uncertainty) in rms spectra \\ 
	14	& MGII$\_$FWHM$\_$mean  & DOUBLE[2] & km s$^{-1}$ &  \MgII\ FWHM (and measurement uncertainty) in mean spectra\\	
	15	& MGII$\_$SIGMA$\_$mean & DOUBLE[2] & km s$^{-1}$ &  \MgII\ $\sigma_{\rm line}$ (and measurement uncertainty) in mean spectra \\
	16	& MGII$\_$FWHM$\_$rms & DOUBLE[2] & km s$^{-1}$ &  \MgII\ FWHM (and measurement uncertainty) in rms spectra \\
	17	& MGII$\_$SIGMA$\_$rms & DOUBLE[2] & km s$^{-1}$ & \MgII\ $\sigma_{\rm line}$ (and measurement uncertainty) in rms spectra \\ 	
	18	& R$\_$FE & DOUBLE & - & Optical Fe II strength (relative to broad \hbeta) \\
	19	& REDCHI2$\_$CONTINUUMFIT & DOUBLE[3] & - & Reduced $\chi^2$ of the continuum fit around \halpha, \hbeta\ and \MgII \\
	20	& REDCHI2$\_$LINEFIT & DOUBLE[3] & - & Reduced $\chi^2$ of line fit of \halpha, \hbeta\ and \MgII \\ 
	21	& FLAG$\_$GOOD$\_$MEAN$\_$HALPHA & INT & -  & {value 1 (0) indicates in (not in) the  \halpha\ good mean subsample}\\
	22	& FLAG$\_$GOOD$\_$RMS$\_$HALPHA  & INT & -  & value 1 (0) indicates in (not in) the \halpha\ good rms  subsample \\
	23	& FLAG$\_$GOOD$\_$MEAN$\_$HBETA & INT & -  & value 1 (0) indicates in (not in) the \hbeta\ good mean subsample  \\
	24	& FLAG$\_$GOOD$\_$RMS$\_$HBETA & INT & -  & value 1 (0) indicates in (not in) the \hbeta\ good rms subsample \\
	25	& FLAG$\_$GOOD$\_$MEAN$\_$MGII & INT & -  & value 1 (0) indicates in (not in) the \MgII\ good mean subsample \\
	26	& FLAG$\_$GOOD$\_$RMS$\_$MGII  & INT & -  & value 1 (0) indicates in (not in) the \MgII\ good rms subsample  \\

         \hline 
        \end{tabular} 

\end{center}
\end{table*}

\section{Line Widths comparison}\label{sec:results}

\subsection{Mean and rms widths for the same line}\label{sec:res1}

\begin{figure}[htbp] 
\begin{center}
\includegraphics[width=0.5\textwidth]{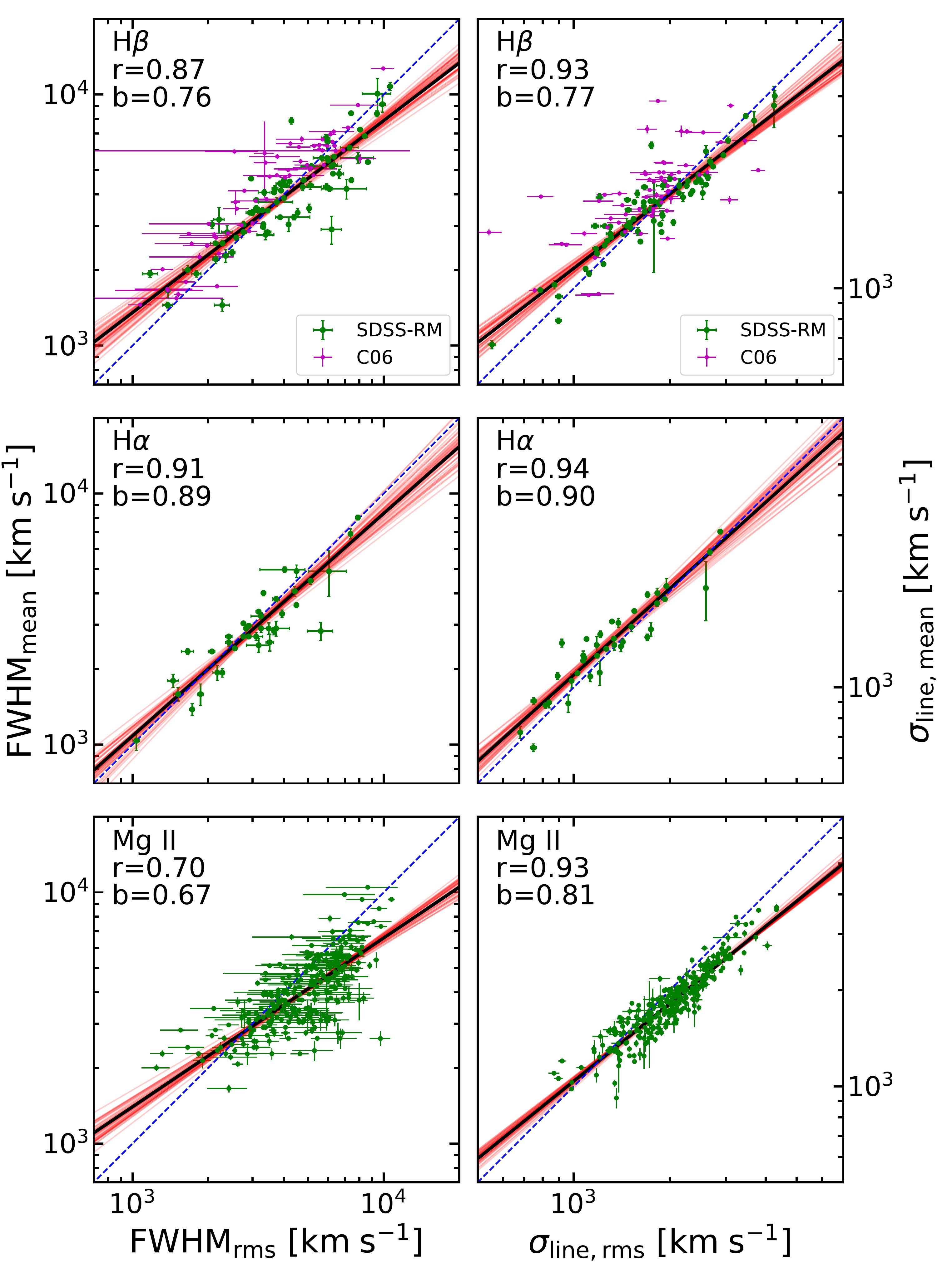} 
\caption{Comparison between FWHM in the mean and rms spectra (left three panels),  and $\sigma_{\rm line}$ in the mean and rms spectra (right three panels) for \hbeta, \halpha\ and \MgII\ from top to bottom, respectively. Green points with error bars are the SDSS-RM sample while magenta points represent the C06 \hbeta-only sample. The blue dashed line in each panel is the 1:1 line. For each comparison, we perform the Bayesian linear regression on the data following \citet{kelly_aspects_2007}. Black solid lines are the median relations from the Bayesian fit and the red shaded regions are the 1$\sigma$ confidence ranges of the fit. The Pearson correlation coefficient $r$ and the best-fit slope $b$ from the linear regression are shown in the upper-left corner in each panel.} \label{fig:correlation} 
\end{center}
\end{figure} 

In this section we use the good mean and rms subsamples to investigate how the mean and rms line widths correlate for the same line.  Apart from SDSS-RM sample, we also include 35 objects  from C06 with mean and rms width measurements of \hbeta. We use the Pearson correlation test to evaluate the significance of each correlation. The Pearson coefficient $r$ reflects the correlation between two quantities, i.e., a value close to 1 indicates that they are tightly correlated. We list the values of $r$, for both FWHM and $\sigma_{\rm line}$ and for \hbeta\ , \halpha\ and \MgII\ in Table \ref{table:statisticcorrelation} and denote them in the upper-left corner of each panel in Figure \ref{fig:correlation}. $r$ for the C06 \hbeta\ sample are also listed in Table \ref{table:statisticcorrelation} for comparison. The slopes of these correlations are measured using the Bayesian linear regression method of \cite{kelly_aspects_2007}. We fit in log-log space using the following equation:

\begin{equation} \label{eqcom0}
\log \Big(W_{\rm mean} / W_0 \Big)= a + b\;  \log \Big(W_{\rm rms}  / W_0 \Big) + \epsilon_0
\end{equation}

\noindent where $a$, $b$ and $\epsilon_0$ are the intercept, slope and intrinsic scatter, respectively. $W_{\rm mean}$ and $W_{\rm rms}$ refer to FWHM or $\sigma_{\rm line}$ from the mean and rms spectrum. $W_0$ is the reference point of the regression fit, whose value is set to 4000 km s$^{-1}$ for FWHM and 2000 km s$^{-1}$ for $\sigma_{\rm line}$, respectively, for both mean and rms line widths. The best-fit parameters, their errors, the intrinsic scatters are listed in Table \ref{table:statisticcorrelation}.

\begin{table*}
\caption {Statistics of mean and rms line width correlations for the same line} \label{table:statisticcorrelation}
\begin{center}
\begin{tabular}{c c c  c  c c c} 
\hline \hline
Line                  				        & Line width definition     & $a$                       &  $b$                 & $\epsilon_0$   & $r$ &  Sample\\ \hline
\multirow{2}{0.5cm}{\hbeta}                  & FWHM              & 0.05 $\pm$ 0.01 & 0.93 $\pm$ 0.04  & 0.05 $\pm$ 0.01  & 0.91 & C06 \\  
 							& $\sigma_{\rm line}$   & 0.04 $\pm$ 0.01 & 0.58 $\pm$ 0.09  & 0.09 $\pm$ 0.01 & 0.63 & C06  \\  

 \multirow{2}{0.5cm}{\hbeta}                  & FWHM             & -0.01 $\pm$ 0.01 & 0.76 $\pm$ 0.05 & 0.08 $\pm$ 0.01 & 0.87 & SDSS-RM\\  
 							& $\sigma_{\rm line}$   & -0.01 $\pm$ 0.01 & 0.77 $\pm$ 0.04 & 0.05 $\pm$ 0.01  & 0.93 & SDSS-RM\\  

 \multirow{2}{0.5cm}{\halpha} 		& FWHM                 & -0.03 $\pm$ 0.01 & 0.88 $\pm$ 0.07 & 0.07 $\pm$ 0.01  & 0.91 & SDSS-RM\\ 
							& $\sigma_{\rm line} $  & 0.01 $\pm$ 0.01 & 0.90 $\pm$ 0.06 & 0.05 $\pm$ 0.01   & 0.94 & SDSS-RM \\  
							
 \multirow{2}{0.5cm}{\MgII}		& FWHM                 & -0.05 $\pm$ 0.01 & 0.67 $\pm$ 0.04 & 0.09 $\pm$ 0.01 & 0.70 & SDSS-RM\\ 
							& $\sigma_{\rm line}$    & -0.04 $\pm$ 0.01 & 0.81 $\pm$ 0.02 & 0.04 $\pm$ 0.01  & 0.93 & SDSS-RM \\  \hline
 \multicolumn{7}{p{0.7\textwidth}}{Notes. Columns 1 and 2 show the line name and line width definition choice studied in each row, respectively. Columns 3, 4 and 5 give the intercept $a$, slope $b$ and intrinsic scatter $\epsilon_{0}$ (in units of dex) from our Bayesian linear regression fitting between mean and rms line widths (see \S\ref{sec:res1}), respectively. Column 6 gives the Pearson correlation coefficient $r$. The last column denotes the sample used to derive those parameters.}
\end{tabular}
\end{center}
\end{table*}

Figure \ref{fig:correlation} compares mean and rms widths for \hbeta, \halpha\ and \MgII\ in each row, respectively. It shows that in general there are strong correlations between the mean and rms widths for each line.  \halpha\ shows slopes close to 1 within 2$\sigma$; \hbeta\ shows slopes slightly shallower but is consistent with a linear correlation. The only exception is the correlation between FWHM$_{\rm mean}$ and FWHM$_{\rm rms}$ of \MgII. It shows a mildly larger scatter and a more non-linear slope than those of the two Balmer lines. This difference is at least partly due to the difficulty of modeling the \MgII\ line in the presence of strong UV \FeII\ emission and our neglect of the fact that \MgII\ is a doublet. The choice of FeII template may also make a difference \citep{Ho_12_UVprofile}. Nevertheless, the strong correlation between \MgII\ $\sigma_{\rm line,mean}$ and $\sigma_{\rm line,rms}$ indicates $\sigma_{\rm line}$ is well correlated between the mean and the variable component of the line; therefore $\sigma_{\rm line,mean}$ could be a better single-epoch virial velocity estimator than FWHM$\rm{_{mean}}$ for \MgII.

C06 reported that for \hbeta, the rms line widths are typically $\sim 20\%$ lower than the mean line widths. As demonstrated by \citet{barth_lick_2015}, it can be affected by the method of constructing the rms spectrum.. The line-only rms spectra generated by \citet{barth_lick_2015} and PrepSpec better represent the variability in the broad lines, and should provide more reliable rms line widths; this improvement probably explains why we find little systematic offset between the mean and rms line widths displayed in Figure\ \ref{fig:correlation}. Nevertheless, consistent with C06, we find a strong correlation between the mean and rms widths for \hbeta.

The correlations between the rms widths and the mean widths provide important justification for single-epoch BH mass recipes, where the widths measured from the single-epoch spectrum are used as a surrogate for the virial velocity of the BLR.

\subsection{Mean and rms widths between different lines}\label{sec:res2}

We now compare the line widths between different broad lines. We adopt \hbeta\ as the reference line in this comparison, since it is the primary line for most RM work in the past and is most commonly used for BH mass estimation. In each comparison we use the same width definition for both lines. The comparisons for the four different width definitions are shown in Figure \ref{fig:hahb} (\halpha\ versus \hbeta) and Figure \ref{fig:hbmg} (\MgII\ versus \hbeta). Only objects included in the good subsamples are used in this comparison. 

As with our analysis in \S\ref{sec:res1}, we perform the Pearson correlation test as well as the Bayesian linear regression on the comparisons between different lines, and the results are summarized in Table \ref{table:hahbmg}. Consistent with earlier work \citep[e.g.,][]{greene_estimating_2005, shen_biases_2008, shen_catalog_2011,shen_comparing_2012,wang_estimating_2009}, there are reasonably good correlations between the broad line widths among the two Balmer lines and \MgII. 

\halpha\ and \hbeta\ are strongly correlated, with slopes close to unity and low scatter. It indicates that \halpha\ widths can be used to substitute for \hbeta\ widths, once the slight difference between the two lines is taken into account. We confirmed that the correlation between \MgII\ and \hbeta\ widths is sub-linear using mean line widths \citep{wang_estimating_2009}. The correlation between \MgII\ and \hbeta\ rms line widths is much stronger, suggesting that the variable component of \MgII\ may also contain information about the virial velocity of the BLR, just as is the case for the Balmer lines.

All previous comparisons between two different lines in earlier work were for the widths measured from the mean spectra, and our study here presents the first systematic cross-line comparison for the widths measured from the rms spectra. The correlations between the widths for different lines provide the justification for using \halpha\ or \MgII\ as an alternative to \hbeta\ for virial BH mass estimation. 

\begin{figure}[h]
\begin{center}
\includegraphics[width=0.5\textwidth]{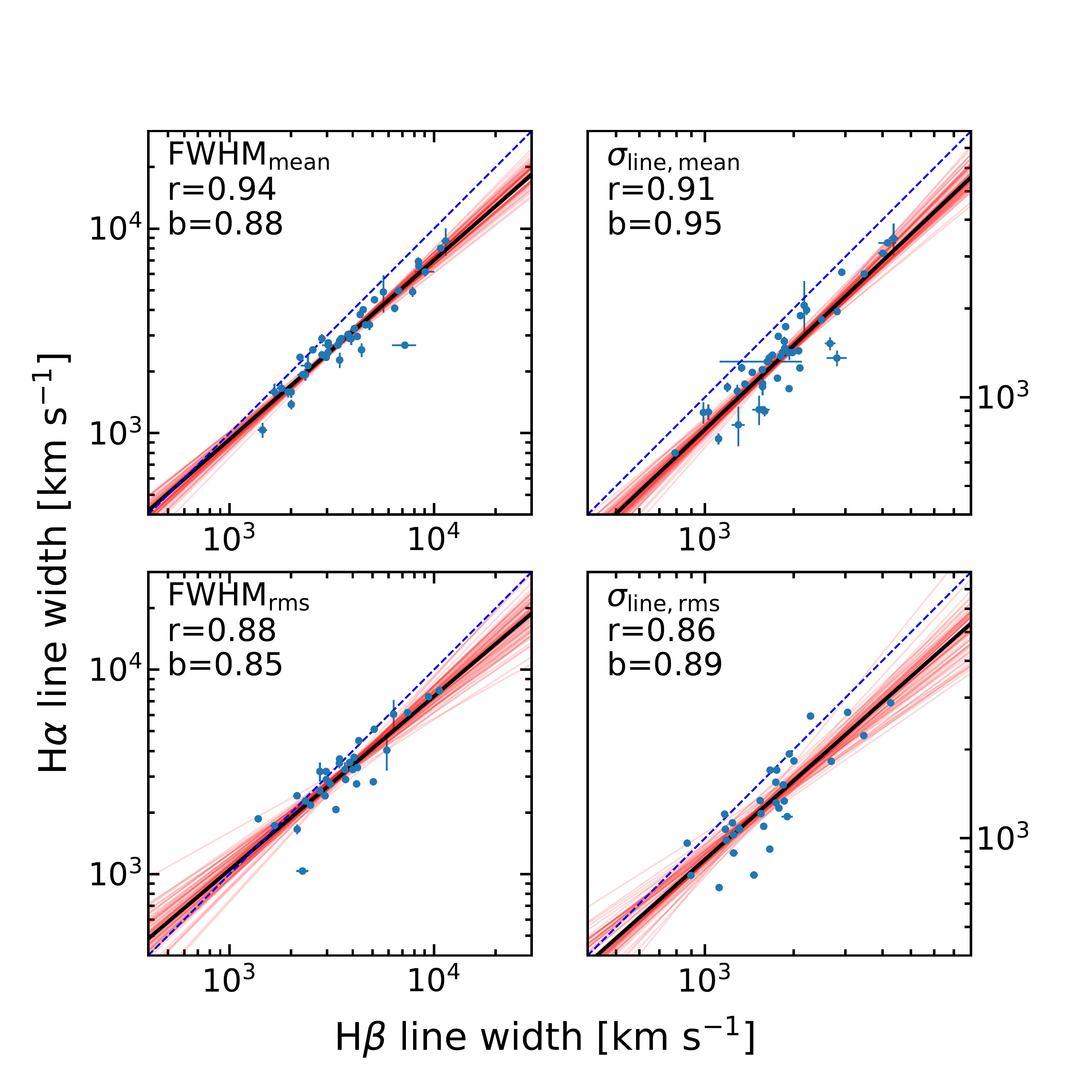}
\caption{Comparison of  the line widths  between \halpha\ and \hbeta\ using four different line width definitions. Blue points with error bars in the upper two panels are objects selected from the good mean \hbeta\ and good mean  \halpha\ subsamples. Blue points with error bars in the bottom two panels represent objects selected from the good rms \hbeta\ and good rms \halpha\ subsamples. Blue dashed lines are the 1:1 correlation. For each line width definition, we perform the Bayesian linear regression fit on the data.  Black solid lines are the median relation from the Bayesian fit and the red shaded regions are the 1$\sigma$ confidence range of the fit. The Pearson correlation coefficients $r$ and slopes $b$ are shown on the upper left corner of each panel.} \label{fig:hahb}
\label{Fig:both}
\end{center}
\end{figure}

\begin{figure}[h]
\begin{center}
\includegraphics[width=0.5\textwidth]{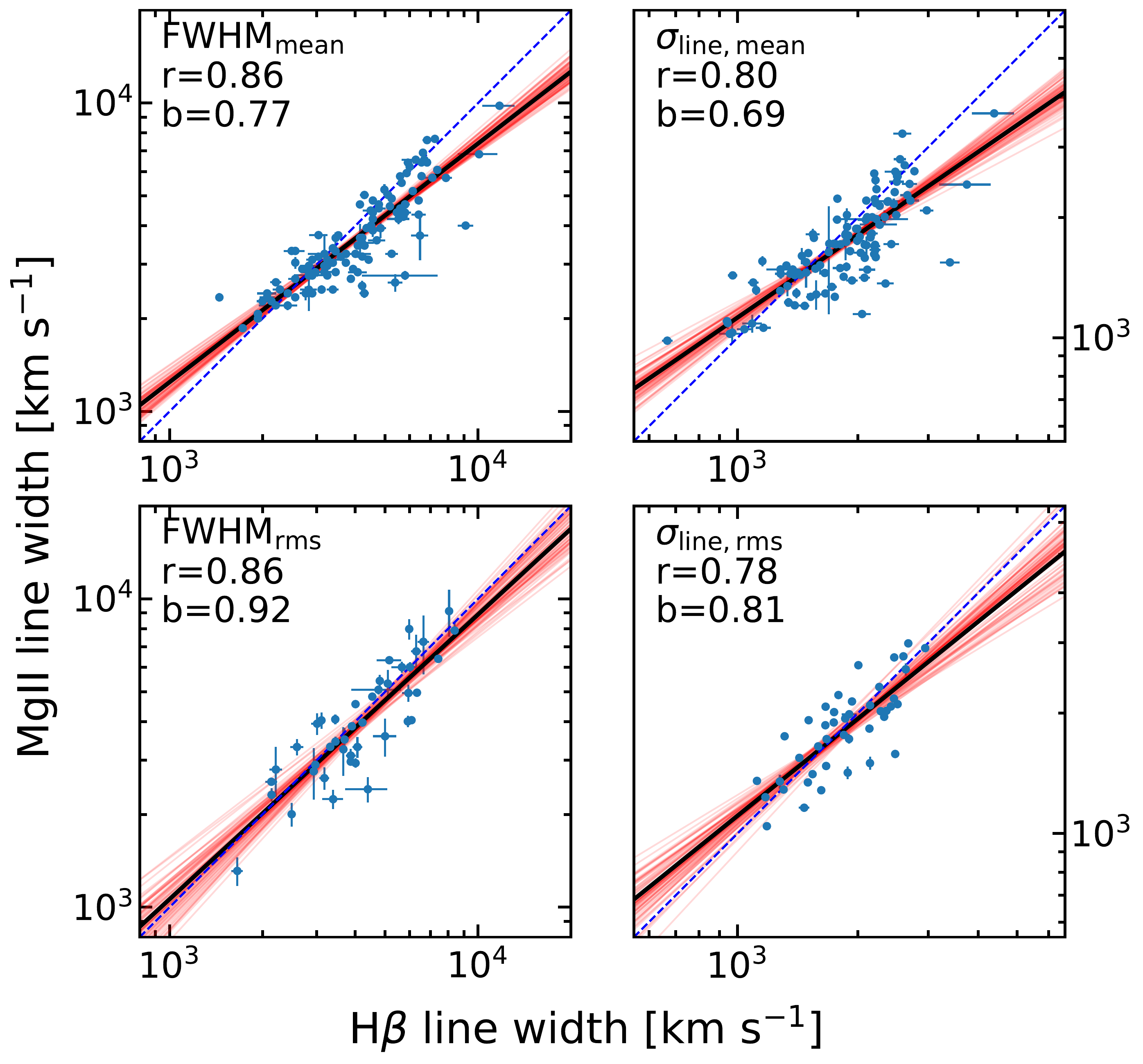}
\caption{Same format as Figure \ref{fig:hahb} but for comparison of \MgII\ and \hbeta. Blue points with error bars in the top two panels represent objects selected from the good mean \hbeta\ and good mean \MgII\ subsample.  Blue points with error bars in the bottom two panels represent objects selected as good rms \hbeta\ and good rms \MgII\ subsample. } \label{fig:hbmg}
\label{Fig:both}
\end{center}
\end{figure}

\begin{table*}
\caption {Statistics of line width comparison between different lines} \label{table:hahbmg}
\begin{center}
 
        \begin{tabular}{c c  c c  c c } 
        \hline  \hline
                                                                             & line width definition & $a$ & $b$  & $\epsilon_{0}$  & $r$ \\ \hline
          \multirow{4}{1.6cm}{\halpha\ vs \hbeta}      & FWHM$_{\rm mean}$           & -0.10 $\pm$ 0.01 & 0.88 $\pm$ 0.05 & 0.05 $\pm$ 0.01 & 0.94 \\ 
                								& $\sigma_{\rm line, mean}$     & -0.12 $\pm$ 0.01 & 0.95 $\pm$ 0.07 & 0.07 $\pm$ 0.01 & 0.91  \\ 
                								& FWHM$_{\rm rms}$                 & -0.07 $\pm$ 0.02 & 0.84 $\pm$ 0.10 & 0.10 $\pm$ 0.01 & 0.88  \\ 
									& $\sigma_{\rm line, rms}$         & -0.11 $\pm$ 0.02 & 0.89 $\pm$ 0.09 & 0.09 $\pm$ 0.01 & 0.86  \\ \hline
									
          \multirow{4}{1.6cm}{\MgII\ vs \hbeta}      & FWHM$_{\rm mean}$               & -0.04 $\pm$ 0.01 & 0.77 $\pm$ 0.04 & 0.07 $\pm$ 0.01 & 0.86 \\ 
                								& $\sigma_{\rm line, mean}$      & -0.04 $\pm$ 0.01 & 0.69 $\pm$ 0.05 & 0.07 $\pm$ 0.01 & 0.80 \\ 
                								& FWHM$_{\rm rms}$                & -0.02 $\pm$ 0.01 & 0.92 $\pm$ 0.09 & 0.09 $\pm$ 0.01 & 0.86 \\ 
									& $\sigma_{\rm line, rms}$       & -0.01 $\pm$ 0.01 & 0.81 $\pm$ 0.11 & 0.07 $\pm$ 0.01  & 0.78  \\ 
\hline 
 
\multicolumn{6}{p{0.6\textwidth}}{Notes. We use the same notation as in Table \ref{table:statisticcorrelation}.}

\end{tabular} 

\end{center}
\end{table*}

\section{Discussion}\label{sec:disc}

\subsection{Which line width is a better virial velocity indicator?}\label{sec:disc1}

To evaluate which line width best indicates the BLR virial velocity and therefore produces the most reliable BH mass estimation, we perform the same test in \citet{collin_systematic_2006}. We assume that the BH masses independently estimated from the $M_{\rm BH}-\sigma_*$ relation, $M_{\rm BH,\sigma_*}$, are reliable and unbiased and we use them as the reference (but see caveats discussed later in \S\ref{sec:disc1}). The virial coefficient $f$ can be computed via the following equation:

\begin{equation}\label{equ:virialfactor}
M_{\rm BH, \sigma_*}  = f \times \frac{W^2\, (c\,\tau) } {G},
\end{equation}

\noindent where $W$ is the broad-line velocity width, either FWHM or $\sigma_{\rm line}$; $c\,\tau$ reflects the average distance of the BLR from the BH; $G$ is the gravitational constant.

Under the assumption that active and inactive galaxies share the same M$_{\rm BH}$-$\sigma_{*}$ relation, \cite{onken_supermassive_2004-1} used 14 RM AGN with host $\sigma_*$ measurements and obtained an average virial coefficient $\langle f \rangle$ = 5.5 $\pm$ 1.8 {using $\sigma_{\rm line,rms}$}. {\cite{Woo_relations_2010, Woo_relations_2013} obtained $\langle f \rangle$ = $5.1^{+1.5}_{-1.1}$ if using only active galaxies and \cite{Grier_stellarvelocity_2013} obtained $\langle f \rangle$ = 4.31$\pm$1.05; both were consistent with \cite{onken_supermassive_2004-1}. However, \cite{Graham_calibration_2011} obtained a virial coefficient of $\langle f \rangle$ = $2.8^{+0.7}_{-0.5}$, which is only half of those reported in previous work.

%However, all these virial coefficients are based on $\sigma_{line, rms}$ given the suggestion from earlier work like C06 that $\sigma_{line, rms}$ better trace the virial velocity. 

In order to address the question of which line width definition is a better indicator for the virial velocity, we compare the virial products (VP) based on each line width definition with the host stellar velocity dispersion, and investigate which line width produces the lowest scatter and least variance in $f$ among different subsets of the sample. We adopt \hbeta\ for this study throughout this section, since most objects with $\sigma_*$ measurements have only \hbeta\ coverage. 

\citet[][hereafter, HK14]{HoKim2014}  compiled all available local RM AGNs with $\sigma_{*}$ measurements and further divided them into subsets of classical bulges and pseudo-bulges. Their study consisted of  35 objects with all four line width definitions, lags and $\sigma_{*}$ measurements available from the literature (see Table 2 in HK14). In SDSS-RM, there are currently 20 objects with all measurements available from the first-year data. For these objects, their $\sigma_{*}$ measurements are taken from \cite{Shen_etal_2015b} where successful $\sigma_{*}$ measurements meet the criteria that they are measured at $>3\sigma$ confidence level, their $\sigma_{*}$ error warning flag is set to zero and their host galaxy fraction is larger than 0.05. Their \hbeta\ lag measurements are taken from \cite{Grier_RM_2017} where successful lag measurements mean that they are at $>2\sigma$ significance.  We cross-match them with our good mean and rms subsample and obtain our final sample. Combined with the HK14 objects, our joint sample includes 55 objects with all quantities available. }

Following HK14, we express the relation between VP and $\sigma_*$ for our quasar sample as
\begin{equation}\label{equ:mcmc}
\log \left(\frac{\rm VP} {M_{\odot}}\right) =  \alpha - \log\; f  + \beta\; \log \left( \frac{\sigma_{*} }  {200\; {\rm km\;s^{-1}}} \right),
\end{equation}

\noindent where $\alpha$ is the normalization, $\beta$ the slope of the $M-\sigma_{*}$ relation, and ${\rm VP}\equiv W^2c\tau/G$ is the virial product from RM observations. We use Markov chain Monte Carlo (MCMC) to perform a linear regression to minimize the quantity:

\begin{equation}\label{equ:chi2}
\chi^2 = \sum_{i=1}^{N} \frac{(y_i + log f - \alpha - \beta x_i)^2} {\epsilon_{yi}^2 + \beta^2\epsilon_{xi}^2+\epsilon_0^2}.
\end{equation}

\begin{figure*}
\begin{center}
\includegraphics[width=0.7\textwidth]{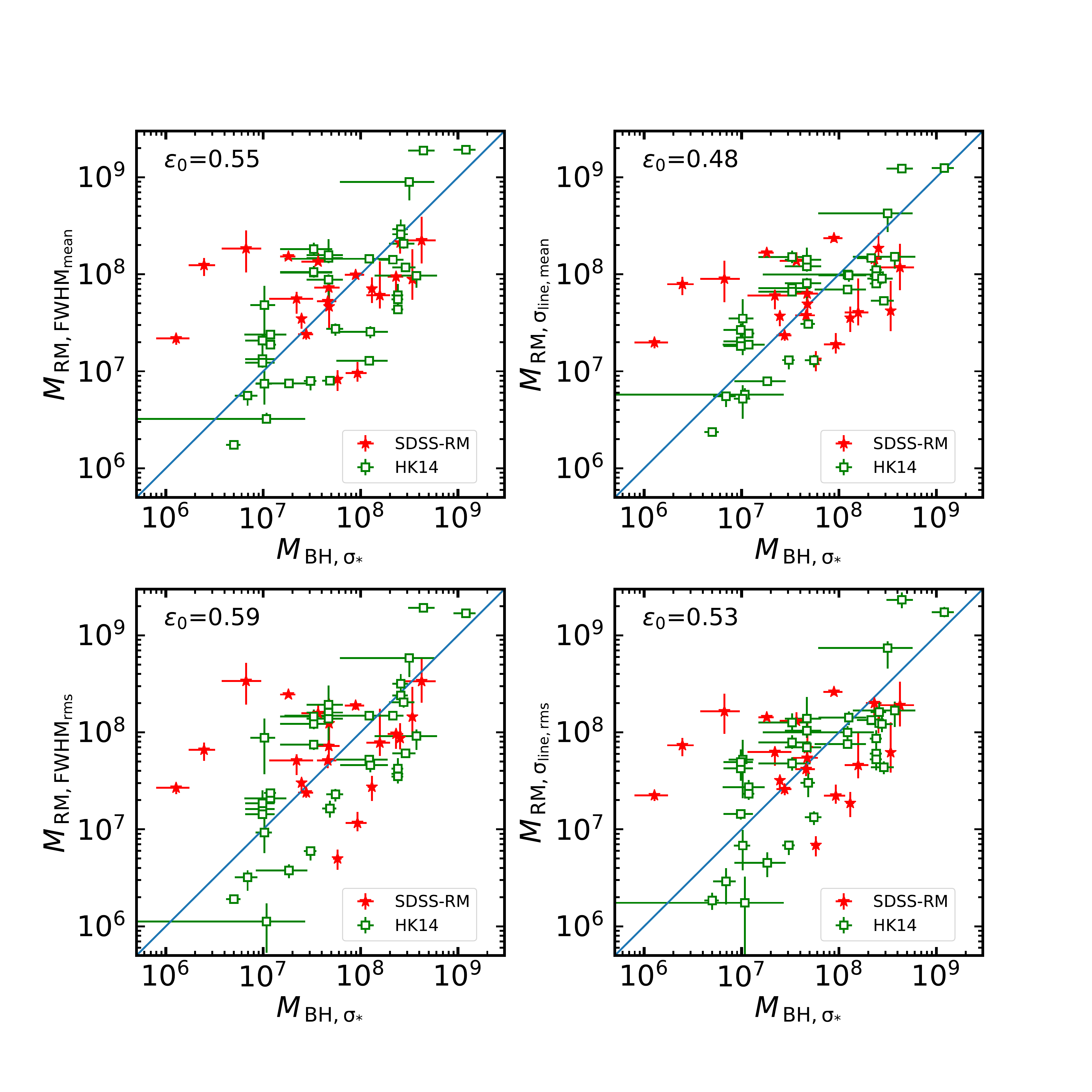}
\caption{Comparison between RM masses derived by multiplying the virial products with an averaged virial coefficient and masses from the M$_{\rm BH}$-$\sigma_{*}$ relation. The four panels are results using four different line width definitions respectively. The intrinsic scatter is labeled in the top-left corner. Red stars denote SDSS-RM objects and green open squares are the HK14 sample. Blue solid lines represent the 1:1 correlation.} 
\label{fig:virialproduct}
\end{center}
\end{figure*}

In the regression fit we fix $\alpha=8.49$ and $\beta=4.38$, from the $M_{\rm BH}-\sigma_*$ relation for hosts classified as classical bulges (CB) in \citet[][see their equation 7]{kormendy_coevolution_2013}.\footnote{Since we lack bulge type information for the SDSS-RM sample, we cannot divide the sample into classical and pseudo bulges (PB) as done in HK14. We treat all objects in our sample as classical bulges.} The results are presented in Figure \ref{fig:virialproduct} for the four line width definitions. The intrinsic scatter $\epsilon_0$ is shown in the upper-left corner of each panel and in Table \ref{table:fscatter}. While the result based on $\sigma_{\rm line,mean}$ has the lowest scatter, the scatter is mutually consistent among the results using all four line width definitions, indicating that our current sample size is insufficient to conclude which width definition is best for virial BH mass estimation based on this particular test.

Table \ref{table:fscatter} also summarizes our estimates of the average virial coefficient $\langle f \rangle$ based on different width definitions of \hbeta\ for the full sample of 55 objects. Our best-fit virial coefficients $\langle f \rangle$ using $\sigma_{\rm line, rms}$ are 6.23$\pm$1.15 which are fully consistent within error bars with those reported in \cite{onken_supermassive_2004-1}, \cite{Woo_relations_2013}, and \cite{Grier_stellarvelocity_2013}.

%Instead of $\sigma_{line, rms}$ that is the traditionally choice of virial velocity indicator,  $\sigma_{line, mean}$ gives the smallest intrinsic scatter with the value of 0.48. 

\begin{table}
\caption{Virial coefficient $f$ and intrinsic scatter $\epsilon_0$ in the VP-$\sigma_*$ correlation} \label{table:fscatter}
\begin{center}
\begin{tabular}{lcc} 

\hline \hline
		Width definition         & $f$	& $\epsilon_0$ \\ \hline
                   FWHM$_{\rm mean}$     & $1.19\pm0.22$	& 0.55 $\pm$ 0.06  \\
                   $\sigma_{\rm line, mean}$ & $4.63\pm0.75$	& 0.48 $\pm$ 0.06  \\
                   FWHM$_{\rm rms}$		& $1.53\pm0.30$ & 0.59 $\pm$ 0.07  \\
                   $\sigma_{\rm line, rms}$    &  $6.23\pm1.15$  & 0.54 $\pm$ 0.06  \\ \hline

\end{tabular}
\end{center}
\end{table}

\begin{table}
\caption{Virial coefficients for different sub-populations} \label{table:subsets}
\begin{center}
\begin{tabular}{p{1.3cm} p{1.5cm} p{1.5cm} p{1.5cm} p{1.5cm}} 

\hline \hline
\                                                 & FWHM$_{\rm mean}$   	 & $\sigma_{\rm line, mean}$  & FWHM$_{\rm rms}$  	        & $\sigma_{\rm line, rms}$ \\ \hline
                   
                   Pop1			& 2.21 $\pm$ 0.56  		& 5.33 $\pm$ 1.09		& 2.74 $\pm$ 0.76		& 8.05 $\pm$ 2.01 \\
                   Pop2			& 0.68 $\pm$ 0.16  		& 4.24 $\pm$ 1.13		& 0.95 $\pm$ 0.25		& 5.18 $\pm$ 1.46 \\
                   PopA       		& 1.81 $\pm$ 0.44  		& 4.86 $\pm$ 0.97		& 2.21 $\pm$ 0.59		& 6.99 $\pm$ 1.64 \\ 
		  PopB     			& 0.72 $\pm$ 0.19 		& 4.63 $\pm$ 1.38		& 1.01 $\pm$ 0.31		& 5.77 $\pm$ 1.82 \\
                   PopC       		& 1.44 $\pm$ 0.43  		& 5.72 $\pm$ 1.54		& 1.97 $\pm$ 0.63		& 8.29 $\pm$ 2.44 \\ 
		  PopD     			& 1.03 $\pm$ 0.27 		& 3.73 $\pm$ 0.77		& 1.23 $\pm$ 0.32		& 4.72 $\pm$ 1.10 \\		  
		  All   		        & 1.19 $\pm$ 0.22		& 4.63 $\pm$ 0.75  		& 1.53 $\pm$ 0.30	        & 6.21 $\pm$ 1.13 \\ 
		  HK14 CB                & 1.5 $\pm$ 0.4                 & 6.3 $\pm$ 1.5                & 1.3 $\pm$ 0.4                 & 5.6 $\pm$ 1.3 \\
		  HK14 PB		       & 0.7 $\pm 0.2$                 & 3.2 $\pm$ 0.7                 & 0.5 $\pm$ 0.2                &  1.9 $\pm$ 0.7 \\  \hline
		  
\multicolumn{5}{p{0.48\textwidth}}{Notes. Columns 2 to 5 list the virial coefficients based on each line width definition. Rows 1 to 6 are virial coefficients calculated using different sub-populations and row 7 is for the full joint sample. The values in rows 8 and 9 are virial coefficients of subsamples with CB and PB taken from Table 3 in HK14.} 

\end{tabular}
\end{center}
\end{table}

Another test to evaluate the line width choice is to investigate if the virial coefficient is consistent among different sub-populations of quasars. Following C06, we divide our joint sample into four sub-populations, Pop1/Pop2 and PopA/PopB. Pop1 and Pop2 are divided using equation (\ref{pop1and2}) below: 

\begin{equation}\label{pop1and2}
{\rm FWHM}_{\rm mean} / \sigma_{\rm line,mean} \sim 2,
\end{equation}

\noindent with Pop1 having smaller ratios. We also divide the sample into PopA and PopB according to equation (\ref{pop3and4}),

\begin{equation}\label{pop3and4}
{\rm FWHM}_{\rm mean} \sim 4000\; km\; s^{-1},
\end{equation}

\noindent with PopA having smaller FWHM$_{\rm mean}$. The two criteria are slightly different from C06 who used FWHM$_{\rm mean}$ / $\sigma_{\rm line,mean}$ $\sim$ 2.35 (value of a Gaussian profile) and  {$\sigma_{\rm line,mean}$} $\sim$ 2000 km~s$^{-1}$. However, to make the subsamples comparable in size between Pop1 and Pop2, we choose the division at FWHM$_{\rm mean}$ / $\sigma_{\rm line,mean}$ $\sim$ 2. The other criterion in C06 is equivalent to ours because the mean ratio of FWHM$_{\rm mean}$ / $\sigma_{\rm line,mean}$ is around 2. With our sample division, there are 26 (29) Pop1 (Pop2) objects, and 31 (24) PopA (PopB) objects. 

In addition to using line width to divide the sample, we also use the \hbeta\ line-continuum flux ratio, $f$(\hbeta)  / $f_{\lambda}$(5100), as a proxy for \hbeta\ line strength (direct measurements of \hbeta\ equivalent width are not publicly available for the HK14 sample) to divide the joint sample into PopC/PopD. The division is:

\begin{equation}\label{pop5and6}
f({\rm H}\beta)  / f_{\lambda}(5100) \sim 60\ [{\textrm \AA}],
\end{equation}

\noindent with PopC having smaller ratios. There are 26 objects in PopC and 28 objects in PopD.

Table \ref{table:subsets} lists the average virial coefficients $\langle f \rangle$ of different sub-populations for the four line widths definitions. For both FWHM$_{\rm mean}$ and FWHM$_{\rm rms}$, there is a large difference in the virial coefficient between Pop1/Pop2 and between PopA/PopB. On the other hand, for both $\sigma_{\rm line,mean}$ and $\sigma_{\rm line,rms}$, the virial coefficients are more consistent across the four different sub-populations. The $\sigma_{\rm line, mean}$ provides the most consistent virial coefficient among these four different quasar sub-populations. These findings are consistent with C06. One interpretation is that FWHM is likely affected by additional parameters that do not trace the underlying virial velocity, such as the orientation of a flattened BLR \citep[e.g.,][]{Wills_Browne_1986,collin_systematic_2006,Shen_Ho_2014,Brotherton_etal_2015,Mejia-Restrepo_etal_2018a}, whereas $\sigma_{\rm line}$ is less sensitive to such parameters.

On the other hand, we find that for both FWHM$_{\rm mean}$ and FWHM$_{\rm rms}$, the virial coefficients are consistent across PopC/PopD  while  for both $\sigma_{\rm line,mean}$ and $\sigma_{\rm line,rms}$ there is a slightly larger difference (only marginally significant) across PopC/PopD. Our previous test comparing the correlations between VPs and host $\sigma_*$ also revealed that using $\sigma_{\rm line}$ produces just as much scatter as using FWHM (e.g., Figure\ \ref{fig:virialproduct}). This could mean that additional factors (other than orientation) likely degrade the correlation between $\sigma_{\rm line}$ and the underlying virial velocity, or even introduce some bias in using $\sigma_{\rm line}$. One possibility is that $\sigma_{\rm line}$ could be measuring parts of the velocities that do not tightly correspond to the virial velocity (e.g., outflows or kinematic components in the profile wings that do not reverberate to the ionizing continuum). Alternatively, $\sigma_{\rm line}$ could indeed be a more reliable indicator for the virial velocity, but this fact is clouded by the systematic uncertainties in the measured quantities that led to Figure\ \ref{fig:virialproduct}, the intrinsic scatter in the $M_{\rm BH}-\sigma_*$ relation, and/or the limited sample size, etc. For example, some $\sigma_*$ measurements may be contaminated by a rotational disk and suffer from orientation effects to some degree, which will introduce extra scatter in the correlation tests in Figure\ \ref{fig:virialproduct}. A larger RM sample with more host bulge $\sigma_*$ measurements is needed to further test these scenarios.

\begin{figure}[htbp] 
\centering
\includegraphics[width=0.5\textwidth]{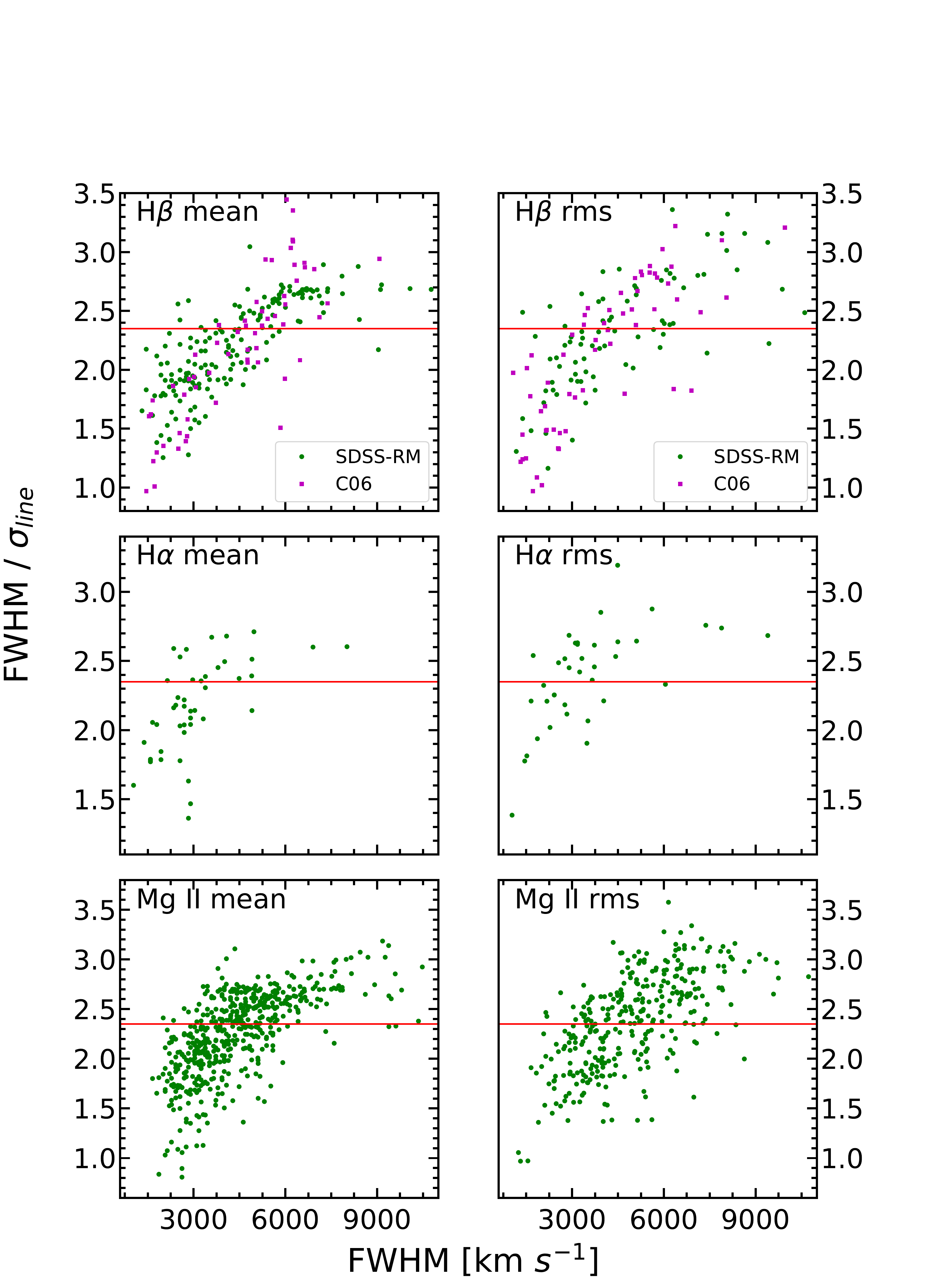} 
\caption{Line shape parameter FWHM/$\sigma_{\rm line}$ as a function of FWHM of mean spectra (left) and rms spectra (right). {The results for \hbeta, \halpha\ and \MgII\ are shown from top to bottom, respectively. } The red horizontal line corresponds to the Gaussian profile with FWHM$/\sigma_{\rm line}=2.35$.} \label{fig:ratiodistribution} 
\end{figure} 

\subsection{Broad line shapes}

Since FWHM is more sensitive to the core of the line while $\sigma_{\rm line}$ depends more on the wings, the ratio FWHM/$\sigma_{\rm line}$ can be used as line shape parameter to characterize the line profile. A Gaussian profile has a FWHM$/\sigma_{\rm line}$ ratio of 2.35; values larger (smaller) than this indicate a higher (lower) fraction of flux in the core than in the wings relative to that for a Gaussian.

Figure \ref{fig:ratiodistribution} demonstrates that the line shape parameter changes as a function of FWHM in mean and rms spectra. There is a trend that the profile becomes more centrally concentrated when FWHM increases. The Pearson correlation coefficients of the trends are presented in Table \ref{table:ratiowidth}. This trend is obvious for both mean and rms profiles and for all three lines. The fact that this trend is not linear (as expected from pure self-correlation due to the common FWHM in both quantities) indicates $\sigma_{\rm line}$ varies in the same direction as FWHM, but with a lower amplitude. 

\begin{table}
\caption{Pearson correlation coefficients $r$ between the ratio FWHM / $\sigma_{\rm line}$ and FWHM \label{table:ratiowidth} }
\begin{center}
%\begin{tabular}{p{2cm}<{\centering} p{1.2cm}<{\centering} p{1.2cm}<{\centering} p{1.2cm}<{\centering} } 
\begin{tabular}{c c c c } 
\hline \hline
		       		 & \hbeta\	& \halpha\ & \MgII\ \\ \hline
        Mean spectra     &  0.75	& 0.59   & 0.65 \\
   	Rms spectra 	 & 0.69	& 0.50  & 0.63 \\
\hline
\end{tabular}
\end{center}
\end{table}

%\blue{In the top two panels of Figure\ \ref{fig:ratiodistribution}, points are color-coded by R$_{Fe}$ where we find there is not a clear trend that this line profile ratio changes with R$_{Fe}$.}

% \halpha\ shows a similar trend as \hbeta\ and \MgII,  although the scatter is larger due to the smaller sample size and the trend seems to flatten when FWHM reaches 5000 km s$^{-1}$.

We further investigate the line profile changing along the Eigenvector 1 (EV1) sequence of quasars. EV1 is a physical sequence that correlates most of the observed quasar properties with the strength of the optical \FeII\ emission \citep[e.g.,][]{boroson_emission-line_1992, Sulentic_etal_2000,Shen_Ho_2014}. In the two dimensional plane of broad \hbeta\ FWHM versus $R_{\rm Fe\,II}\equiv {\rm EW_{Fe\,II,4434-4684} /EW_{H\beta}}$, EV1 is defined as a band extending from low to high $R_{\rm Fe\,II}$ with decreasing average broad \hbeta\ FWHM. \citet{Shen_Ho_2014} suggested that the vertical dispersion in broad \hbeta\ FWHM in the EV1 plane is mainly due to an orientation effect (combined with any intrinsic dispersion of line width due to different BH masses sampled), where the differences in FWHM reflect the changes in the orientation of a flattened BLR along the line-of-sight. If this interpretation is correct, there should be a similar vertical dispersion of the shape parameter within the EV1 sequence, if $\sigma_{\rm line}$ is less susceptible to orientation. 

%\blue{Alternatively, if the vertical dispersion in FWHM in the EV1 plane is mainly due to intrinsic changes in the virial velocity, then $\sigma_{\rm line}$ should follow a similar dispersion and hence the FWHM/$\sigma_{line}$ ratio should remain roughly the same in the vertical direction. } \red{Could you make a plot similar to Figure 7, but color-coded by $\sigma_{line}$? I wonder if $\sigma_*$ is constant vertically. }

Figure \ref{fig:ev1} shows the distribution of the broad \hbeta\ shape parameter in the EV1 plane. We use FWHM and $\sigma_{\rm line}$ from either mean or rms spectra and compute the shape parameter from their ratios. The shape parameter measured from the mean spectrum (left panel) displays a clear vertical segregation at fixed $R_{\rm Fe\,II}$, consistent with the findings in \citet[][see their Figure E10]{Shen_Ho_2014}. A similar trend exists for the shape parameter measured from the rms spectrum (right panel of Figure\ \ref{fig:ev1}). Interestingly, the segregation of objects with different line shapes more or less tracks the main EV1 sequence such that the running median of the distribution at fixed $R_{\rm Fe\,II}$ has roughly the same line shape parameter. This result is fully consistent with the framework discussed in \citet{Shen_Ho_2014}, where the running median of the EV1 sequence represents the average orientation of a flattened BLR, and vertical deviations from the median correspond to variations of the orientation and hence the changes in the line shape parameter.

\begin{figure*} 
\centering
\includegraphics[width=0.7\textwidth]{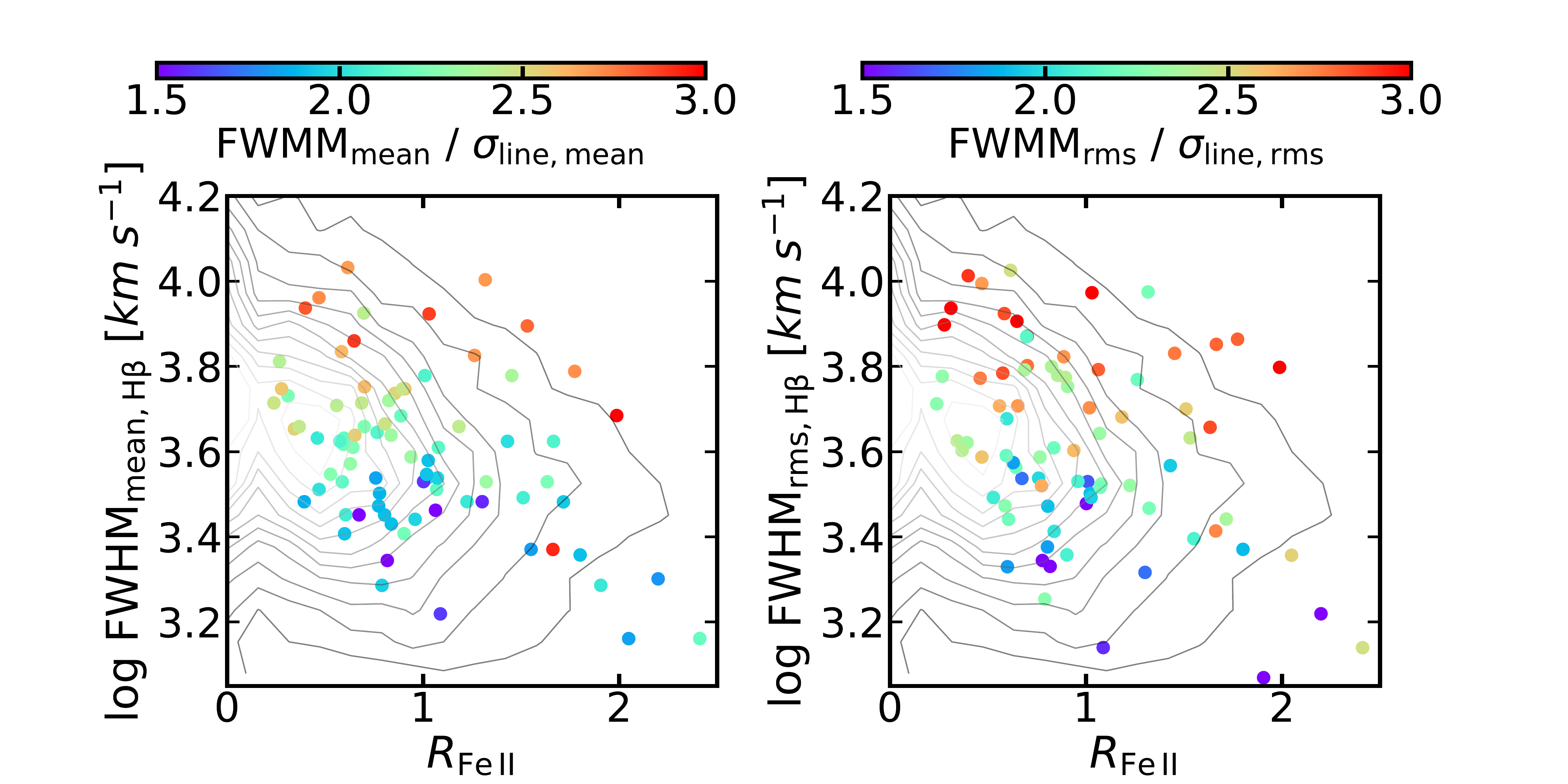} 
\caption{Distribution of the SDSS-RM sample in the EV1 plane color-coded by the line shape parameter FWHM/$\sigma_{\rm line}$, for the mean widths (left) and rms widths (right). In both panels there is a clear segregation of colors that tracks the EV1 sequence. The distributions of the SDSS DR7 quasars from \cite{shen_catalog_2011} are indicated as the gray contours.} \label{fig:ev1}
\end{figure*} 

\section{Conclusions}\label{sec:con}

We have investigated different definitions of line widths from the mean and rms spectra for three low-ionization broad lines, \halpha, \hbeta\ and \MgII, using the sample of quasars from the SDSS-RM project. For each of the three lines, we fit and measured the broad-line FWHM and $\sigma_{\rm line}$ from the mean and rms spectra derived from the first-season spectroscopic observations of SDSS-RM. We compared different definitions of line widths for the same line, and compared the same line widths among different lines.

The main results from this study are:
\begin{itemize} 

\item[$\bullet$] We introduced a new recipe to measure $\sigma_{\rm line}$ that is reproducible, less susceptible to noise and blending in the wings, and scales with the intrinsic width of the line. This quantitative recipe can be used to measure consistent $\sigma_{\rm line}$ values by different groups. 

\item[$\bullet$] There are significant correlations between different line widths for the same line. Since most of the RM BH masses to date are based on the rms $\sigma_{\rm line}$, the correlations between other width definitions and $\sigma_{\rm line}$ provide justification for using the other width definitions for single-epoch virial BH mass estimation where only the mean FWHM and $\sigma_{\rm line}$ are available. 

\item[$\bullet$] There are also strong correlations between the line widths for different lines. The consistency in broad-line widths between the other two lines (\halpha\ and \MgII) and the \hbeta\ line provides the justification for using these alternative lines rather than \hbeta\ to estimate virial BH masses. 

\item[$\bullet$] We investigated the correlations between the RM virial products based on different width definitions and the host stellar velocity dispersion $\sigma_*$, using a sample of 55 quasars with both RM and $\sigma_*$ measurements. We calculated the virial coefficient $f$ using the $\sigma_*$ measurements for different width definitions and for different sub-populations of quasars. Consistent with C06, we found that the virial coefficient using $\sigma_{\rm line}$ is more or less consistent across sub-populations divided by line width, while the virial coefficient using FWHM has a large dispersion among these sub-populations. On the other hand, the virial coefficient using $\sigma_{\rm line}$ is less consistent than that using FWHM across sub-populations divided by line strength, although the significance of the difference is low. The correlation analysis for the virial product versus stellar velocity dispersion for the full sample does not provide conclusive evidence that any of the line width definitions is better than the others. 

\item[$\bullet$] We studied the shape of the broad \hbeta\ line, characterized by FWHM/$\sigma_{\rm line}$, as functions of quasar parameters. Consistent with \citet{Shen_Ho_2014}, we found that the shape parameter displays a segregation in the EV1 plane that closely follows the sequence along EV1, which again suggests that orientation plays a stronger role in changing the FWHM of the line than $\sigma_{\rm line}$.

\end{itemize} 

Our results are consistent with earlier studies \citep[e.g.,][]{greene_estimating_2005,collin_systematic_2006,shen_biases_2008, shen_catalog_2011,shen_comparing_2012, wang_estimating_2009}, but extend to all four line width definitions with a much larger RM sample and for all three low-ionization broad lines. In particular, the comparisons involving the rms widths of \halpha\ and \MgII\ are presented here for the first time. 

Although our results corroborate earlier suggestions that FWHM may be more impacted by orientation effects in a flattened BLR than $\sigma_{\rm line}$ \citep[e.g.,][]{Wills_Browne_1986,Runnoe_etal_2013a,Shen_Ho_2014,Brotherton_etal_2015,Mejia-Restrepo_etal_2018a}, it does not necessarily mean $\sigma_{\rm line}$ is a more accurate tracer of the virial velocity for the broad lines. $\sigma_{\rm line}$ is sensitive to the wings of the line, which may not reverberate to continuum changes and/or may arise in a non-virial component. To determine whether FWHM or $\sigma_{\rm line}$ is a better tracer of the virial velocity, additional data are required for two case studies: (1) a large sample of quasars with both host stellar velocity dispersion and broad-line width measurements to determine which width definition produces the tightest correlation between the virial product and the stellar velocity dispersion; (2) a large sample of quasars with large dynamical range in their continuum variability to test which width definition best follows the expected virial relation \citep[e.g.,][]{peterson_central_2004,shen_mass_2013}. Our attempt on this first case study did not yield conclusive results, but future larger samples with both RM and $\sigma_*$ measurements may clarify the situation. The SDSS-RM project will compile multi-year light curves for 849 quasars and thus will provide one of the best samples for the second investigation. In addition, the multi-year data from SDSS-RM will be used to extend our line width study to other broad lines (such as \CIV) covered in the high-redshift subsample. 

\acknowledgements
We thank Brad Peterson and the anonymous referee for suggestions that improved this work. We acknowledge support from the National Key R\&D Program of China (2016YFA0400703), the National Science Foundation of China (11533001, 11721303, 11890693), and the Chinese Academy of Sciences (CAS) through a China-Chile Joint Research Fund \#1503 administered by the CAS South America Center for Astronomy. YS acknowledges support from an Alfred P. Sloan Research Fellowship and NSF grant AST-1715579. KH acknowledges support from STFC grant ST/R000824/1. CJG acknowledges support from NSF grant AST-1517113. WNB acknowledges support from NSF grants AST-1517113 and AST-1516784.

%WNB and CJG acknowledge support from NSF grant AST-1517113. KDD is supported by an NSF AAPF fellowship awarded under NSF grant AST-1302093. BMP is grateful for support by the NSF through grant AST-1008882 to the Ohio State University. 

Funding for SDSS-III has been provided by the Alfred P. Sloan Foundation, the
Participating Institutions, the National Science Foundation, and the U.S.
Department of Energy Office of Science. The SDSS-III web site is
http://www.sdss3.org/.

SDSS-III is managed by the Astrophysical Research Consortium for the
Participating Institutions of the SDSS-III Collaboration including the
University of Arizona, the Brazilian Participation Group, Brookhaven National
Laboratory, University of Cambridge, Carnegie Mellon University, University
of Florida, the French Participation Group, the German Participation Group,
Harvard University, the Instituto de Astrofisica de Canarias, the Michigan
State/Notre Dame/JINA Participation Group, Johns Hopkins University, Lawrence
Berkeley National Laboratory, Max Planck Institute for Astrophysics, Max
Planck Institute for Extraterrestrial Physics, New Mexico State University,
New York University, Ohio State University, Pennsylvania State University,
University of Portsmouth, Princeton University, the Spanish Participation
Group, University of Tokyo, University of Utah, Vanderbilt University,
University of Virginia, University of Washington, and Yale University.

\newpage
\appendix
\renewcommand\thefigure{\Alph{section}\arabic{figure}}

\section{A. Simulations of the RMS Spectrum Generation}\label{sec:sim}

As mentioned in \S\ref{sec:sample}, the traditional approach of rms spectrum construction includes the variations from the continuum. \cite{barth_lick_2015} reported that the continuum variation and random noise in the spectrum can cause the generated rms line profile to deviate substantially from the true rms line profile, that is, solely due to line variation. They adopted a decomposition approach to isolate the continuum component in individual epochs before the construction of the line-only rms spectrum, similar to our PrepSpec approach. They used simulations to demonstrate the differences in the line rms profile from these two approaches, and below we perform a similar exercise to reproduce their results.  

The numerical experiment is to examine a potential bias in rms line width that occurs when the rms spectrum is constructed from the entire spectrum from individual epochs. To examine this potential bias, we perform a simple simulation to illustrate the difference in the line profile between the two approaches of rms spectrum generation. Using the method of \cite{timmer_generating_1995}, we construct a random light curve with a certain duration, sampling, power spectrum slope and rms variability amplitude $F_{var}$. We assume a fluctuation power spectrum $P(\mu) \sim \mu^{-2.7}$, where $\mu$ is the frequency in Fourier space. The power-law index is similar to the variability properties of nearby AGNs monitored by the Kepler emission \citep{mushotzky_kepler_2011}. A series of single-Gaussian line profiles whose line luminosity and width vary in response to continuum fluctuations is generated. For simplification we assume the integrated luminosity of the line at time $t$ scales linearly with the input continuum light curve luminosity at time $t-\tau$, where $\tau$ is the assumed time lag for the velocity-integrated transfer function. The width of the Gaussian profile also varies accordingly such that the product $\sigma^{2}L^{0.5}$ remains constant when luminosity varies, as expected from a perfect virial relation for the line. The broad  emission line is set to have a mean equivalent width of 120\AA. The initial width is set to be $\sigma_{\rm line}$ = 2000 ${\rm km\,s^{-1}}$ and the shape of the continuum is set to be flat over the wavelength range of interest. Spectra are constructed in velocity bins of 60 ${\rm km\,s}^{-1}$. Then, the two sets of spectra, one with emission line only, the other with the total flux including both line and continuum, are created. The FWHM and $\sigma_{\rm line}$ are measured on both sets of rms spectra. Since the continuum and emission line variations are occasionally out of phase, there could be a pair of local minima in total-flux rms spectrum; in such cases we calculate the $\sigma$ between the two local minima. 

The results of the simulations demonstrate that the  total-flux rms spectrum line width could differ significantly from that of the line-only rms spectrum. Figure\ \ref{Fig:simulationexample} presents an example of our simulations. In this example, FWHM(total-flux) / FWHM(line-only) is 0.766 and $\sigma$(total-flux) / $\sigma$(line-only) is 0.740, indicating that the widths measured from the rms spectrum generated by the traditional approach are biased low by $\sim 20-30\%$ relative to those measured from the true line rms spectrum. 

\setcounter{figure}{0}  
\begin{figure*}
\begin{center}
\begin{minipage}[b]{0.49\textwidth} 
\centering
\includegraphics[width=1.0\textwidth]{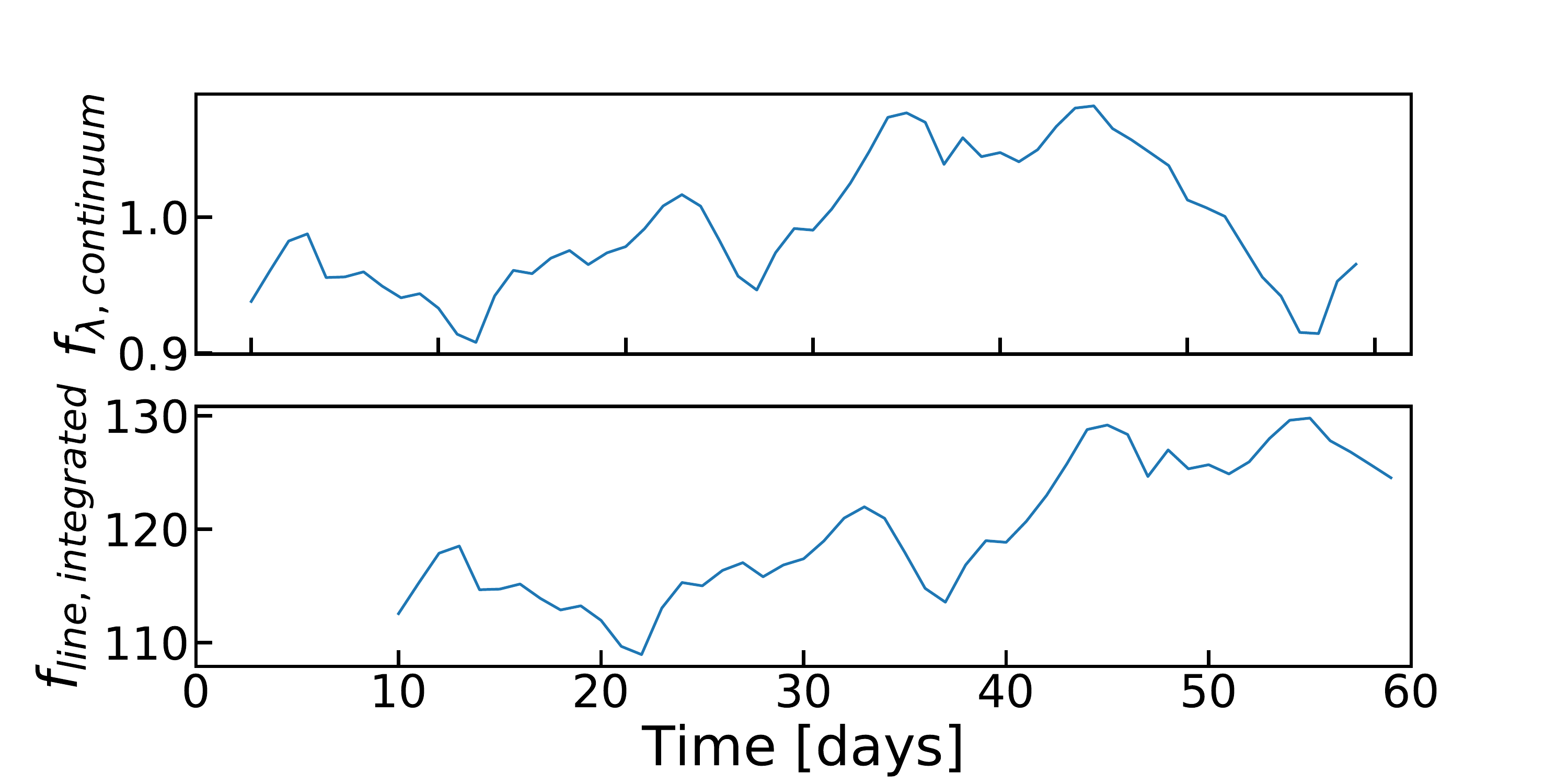}
\includegraphics[width=1.0\textwidth]{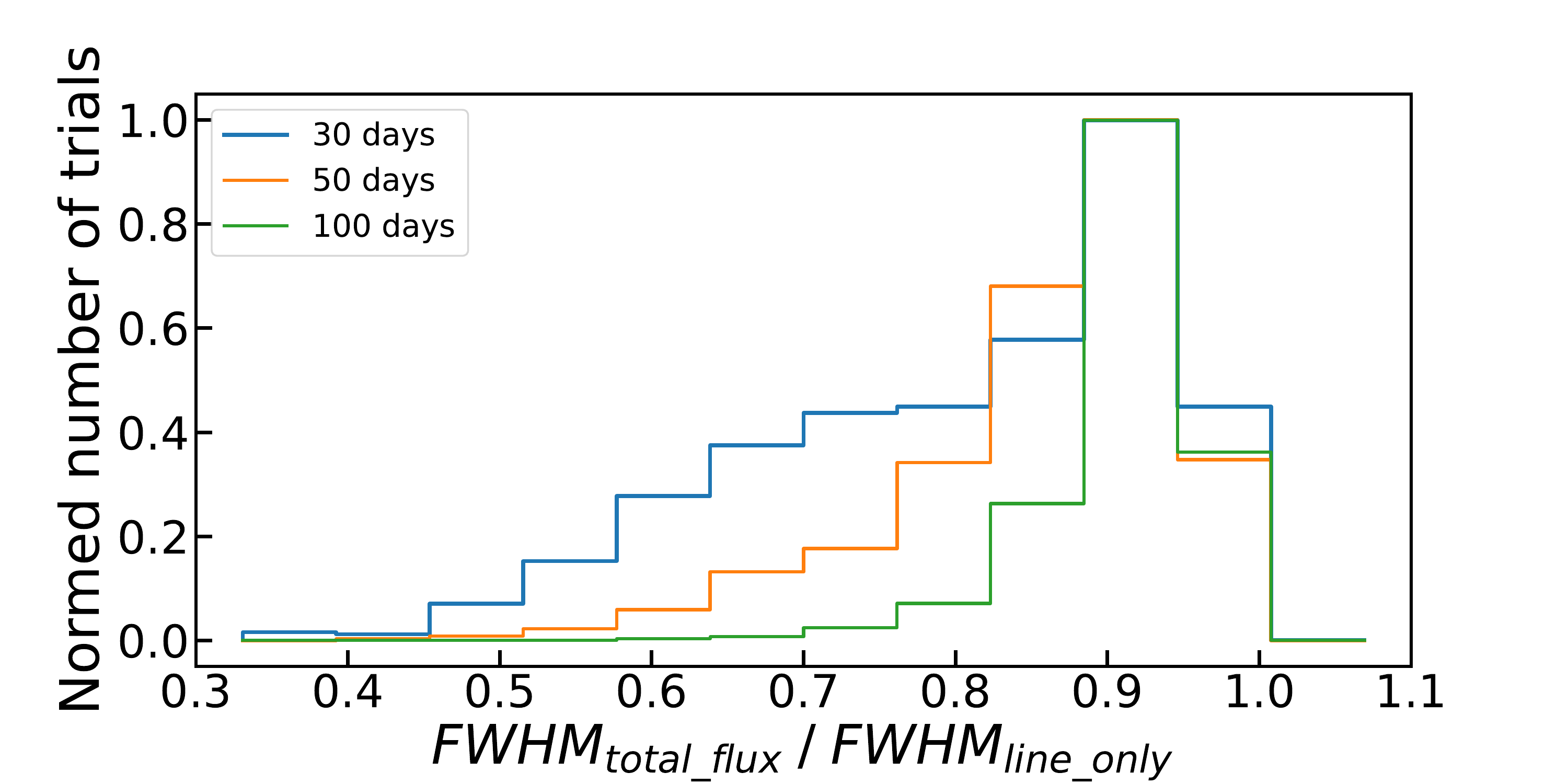}
\end{minipage}% 
\begin{minipage}[b]{0.49\textwidth} 
\centering
\includegraphics[width=1.0\textwidth]{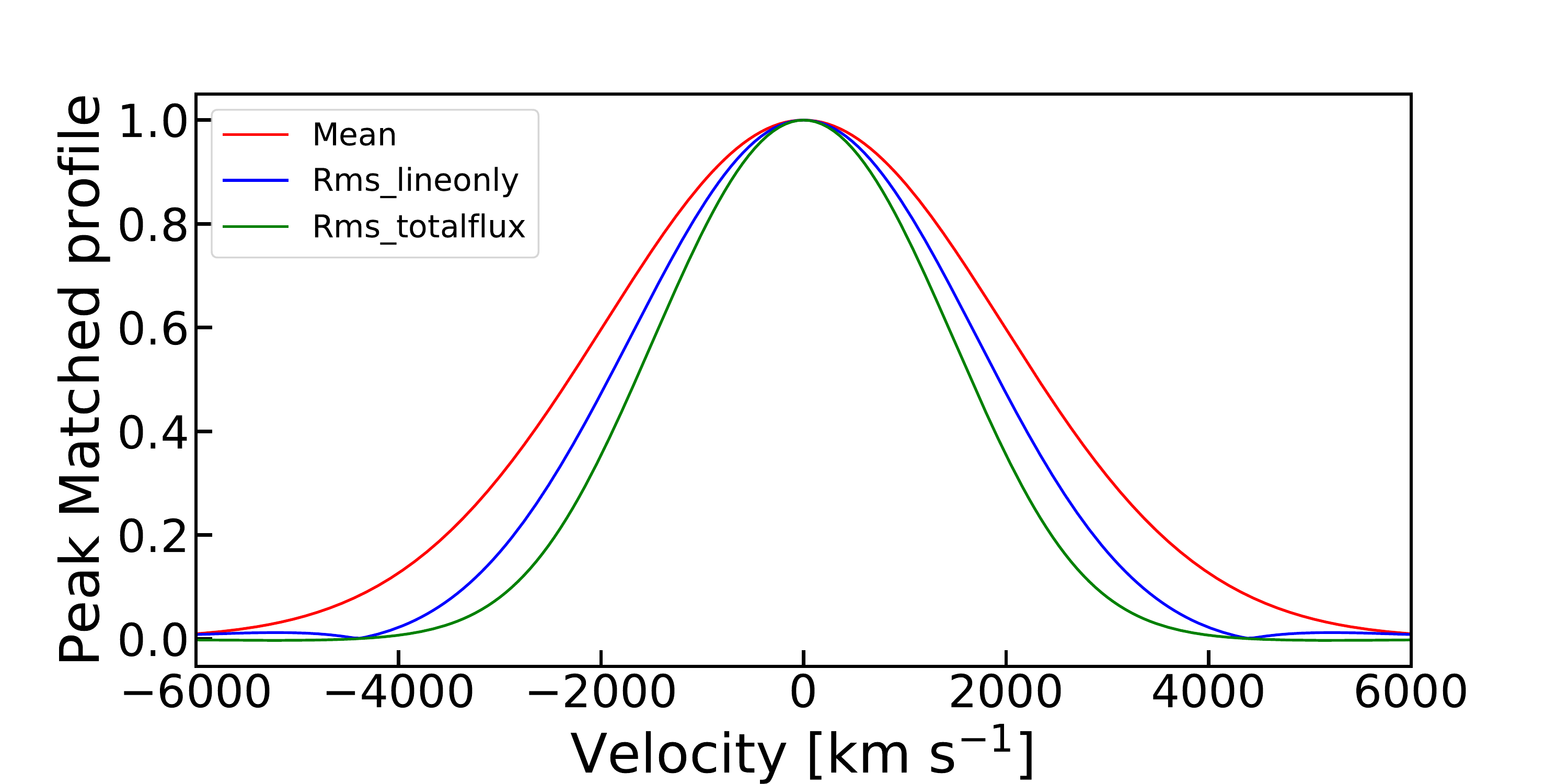}
\includegraphics[width=1.0\textwidth]{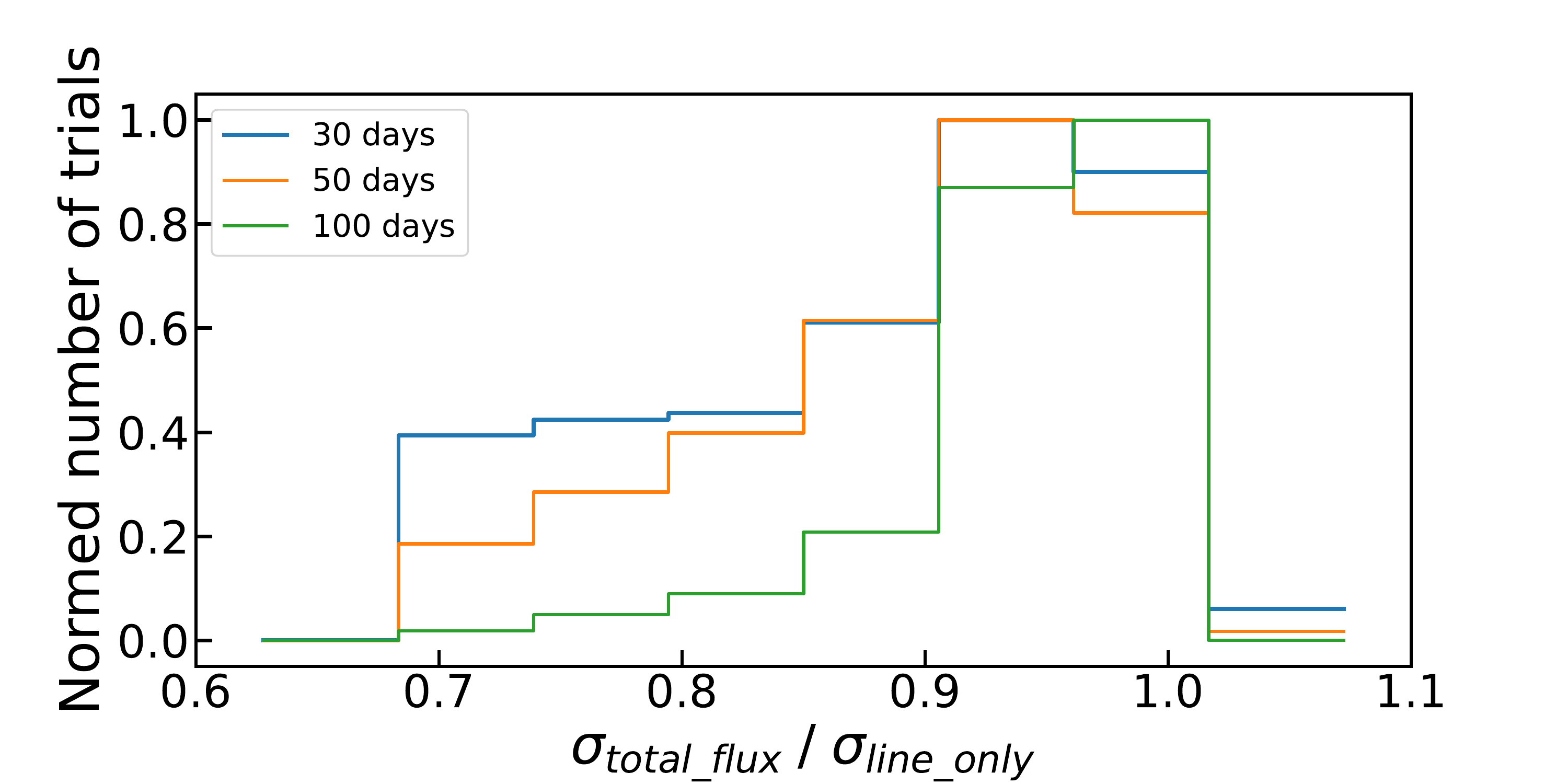}
\end{minipage} 
\caption{Upper-left panel: an example of simulated continuum and emission line light curves. The emission line light curve has a 10-day lag with respect to the continuum light curve. Upper-right: mean spectrum (red), line-only rms spectrum (blue), and total-flux rms spectrum subtracted by a local continuum (green). Lower panels: the distribution of the line width ratios between line width in total$\_$flux rms spectra and line$\_$only rms spectra using FWHM (left) and $\sigma_{\rm line}$ (right), based on 1000 trials for this particular example. The Blue, orange, and green histograms refer to 30, 50 and 100 days of the nightly RM observation baseline. Shorter programs can lead to more significantly biased rms line width.}
\label{Fig:simulationexample}
\end{center}
\end{figure*}

%This bias occurs because of the out-of-phase variations between the continuum and the emission line. The emission line variation responds to continuum variations with a time lag. These variations are stochastic and sometimes out-of-phase with each other. In the line wing region, the rms spectrum is dominated by continuum variation and in line core region, the rms spectrum is dominated by line variation. However, though both of them are stochastic, there will be some part of the light curves where the variation of continuum and emission line happens to be opposite, e.g. the continuum is rising while the emission line flux is falling and vice versa.  This opposite phase variation tends to preferentially suppress the line wing  relative to the line core of rms spectrum from total flux that is not subtracted continuum first when constructing. Consequently the rms profile constructed  from total flux tends to have smaller line width than that constructed from pure emission line spectrum. Even a modest bias in determination of line width can have a big effect on black hole mass because black hole mass scales as the square of the line width.

To quantify the magnitude of this bias, 1000 sets of simulated light curves and spectra with different monitoring durations of 30, 50, and 100 days, assuming nightly sampling, were created. The variability amplitude $F_{var}$ of the continuum light curve is 0.1.  Figure \ref{Fig:simulationexample} shows the bias distribution for both FWHM (lower-left) and $\sigma_{\rm line}$ (lower-right). For both $\sigma_{\rm line}$ and FWHM, the bias in the rms line widths measured from the traditional approach depends on the monitoring duration. For relatively short reverberation mapping programs, the traditional approach can lead to significantly biased rms line width.

\section{B. The impact of windows on the $\sigma_{\rm line}$ calculation}

In the $\sigma_{\rm line}$ calculation the integral kernel is proportional to $\lambda^2$, hence $\sigma_{\rm line}$ is extremely sensitive to the computation window. This behavior is especially true for objects with moderate S/N, where it is difficult to determine the boundaries of the line.
\setcounter{figure}{0}  
\begin{figure*}
\centering
\includegraphics[width=0.7\textwidth]{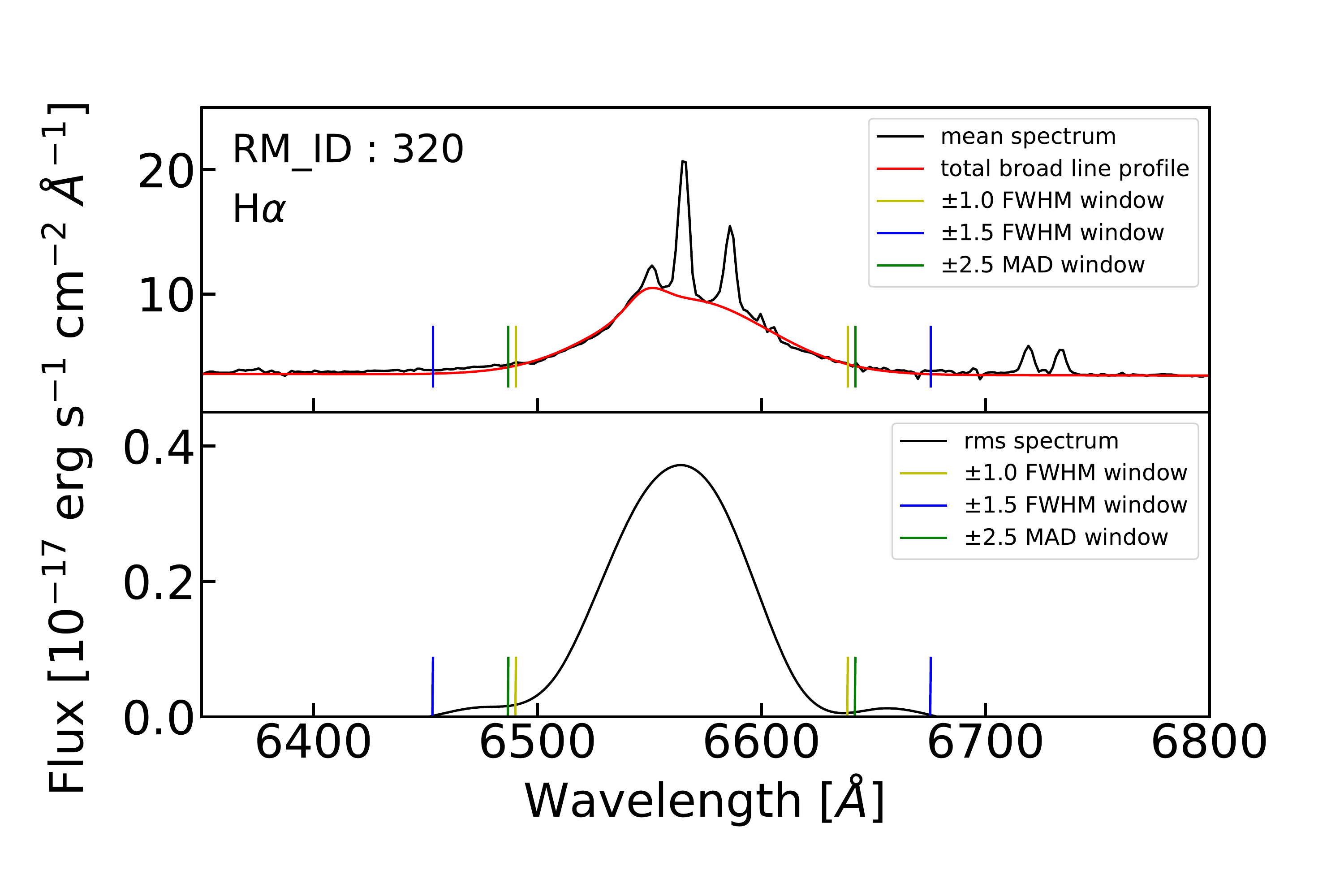}
\caption{Illustration of the three different choices of window size. The black solid lines in the upper and lower panels represent the mean and rms spectra, respectively. The red line is the total broad line profile in the mean spectrum used to calculate different line widths. The yellow, blue and green vertical lines represent 1.0$\times$FWHM window, 1.5$\times$FWHM window and 2.5$\times$MAD window, respectively.}
\label{Fig:choiceofwindow}

\end{figure*}

In this work we defined a quantitative window to compute $\sigma_{\rm line}$. This window is determined from the line width measured in the mean spectrum, which typically has much higher S/N than the rms spectrum. We experimented with three choices to define this window, with the half window size equal to FWHM$_{\rm mean}$, 1.5$\times$FWHM$_{\rm mean}$, and 2.5$\times$MAD$_{\rm mean}$ as shown in Figure \ref{Fig:choiceofwindow}. For mean spectra, the center of the window is chosen as the zero velocity measured from the rest-frame vacuum wavelength. Whereas for rms spectra, the center of the window is chosen to be the peak of the rms profile to better include much of the rms flux, since the rms profile is often highly asymmetric. In this example the peak of rms profile is close to the vacuum wavelength, thus the mean and rms windows are similar.

Based on visual inspection of our objects, the 2.5$\times$MAD$_{\rm mean}$ half window size best balances the need to enclose most of the line flux and to avoid noisy wings in the calculation of $\sigma_{\rm line}$. Since this window size scales with the width of the line (e.g., broader lines will yield larger widow sizes), the $\sigma_{\rm line}$ will not be biased as a function of line width, compared to the case of a fixed window size. Conversely, the other two window sizes are based on FWHM$_{\rm mean}$, which is sensitive to the central part of the profile. Hence, it is not as good as using MAD$_{\rm mean}$, which includes contributions from the wings. Indeed, the tightest correlations are those with $\sigma_{\rm line}$ calculated using our fiducial window defined by MAD$_{\rm mean}$, compared with those calculated using the windows defined by FWHM$_{\rm mean}$.  We therefore adopt $2.5\times$MAD$_{\rm mean}$ as the fiducial window to compute $\sigma_{\rm line}$ for both the mean and rms spectra. Each line has its own MAD measurement and thus different lines in the same object have slightly different window sizes.  

%\newpage
%\begin{thebibliography}{}
\bibliographystyle{apj}
%\bibliography{/Users/wangshu/Documents/work/reverberationmapping/linewidth_v2/ref}
\bibliography{ref}

%\bibliography{refs}
%\end{thebibliography}

\end{document}